\documentclass[prd,showpacs,nofootinbib,preprintnumbers,preprint]{revtex4}
\renewcommand{\theequation}{\arabic{section}.\arabic{equation}}
\usepackage{amsmath}  
\usepackage{amsfonts}
\usepackage{graphicx}
\def\be{\begin{equation}}
\def\ee{\end{equation}}
\def\ba{\begin{eqnarray}}
\def\ea{\end{eqnarray}}
\def\dfrac{\displaystyle\frac}
\def\nn{\nonumber}
\def\lb{\label}
\def\bb{\bibitem}
\def\nn{\nonumber}
\def\r{\hat{r}}

\def\C{{\cal C}}

\def\ol{\overline}

\begin{document}
\title{\begin{flushright}\begin{small}    LAPTH-047/15
\end{small} \end{flushright} \vspace{1.5cm}
NUT wormholes}
\author{G\'erard Cl\'ement} \email{gerard.clement@lapth.cnrs.fr}
\affiliation{LAPTh, Universit\'e Savoie Mont Blanc, CNRS, 9 chemin de Bellevue, \\
BP 110, F-74941 Annecy-le-Vieux cedex, France}
\author{Dmitri Gal'tsov} \email{galtsov@phys.msu.ru}
\affiliation{Department of Theoretical Physics, Faculty of Physics,
Moscow State University, 119899, Moscow, Russia}
\author{Mourad Guenouche} \email{guenouche_mourad@umc.edu.dz}
\affiliation{ Laboratoire de Physique Th\'eorique, D\'epartement de
Physique, \\ Facult\'e des Sciences Exactes,
Universit\'e de Constantine 1, Algeria;  \\
Department of Physics, Faculty of Sciences, Hassiba Benbouali
University of Chlef, Algeria}
\begin{abstract}
We show that supercritically charged black holes with NUT provide a
new setting for traversable wormholes. This does not require exotic
matter, a price being the Misner string singularities. Without
assuming time periodicity to make Misner strings unobservable, we
show that, contrary to expectations, geodesics do not stop there.
Moreover, since there is no central singularity the space-time turns
out to be geodesically complete.  Another unpleasant feature of
spacetimes with NUTs is the presence of regions where the azimuthal
angle $\varphi$ becomes timelike, signalling the appearance of
closed timelike curves (CTCs). We show that among them there are no
closed timelike or null geodesics, so the freely falling observers
should not encounter causality violations. Considering  worldlines
of charged particles, we find that, although these can become closed
in the vicinity of the wormhole throat for large enough
charge-to-mass ratio, the non-causal orbits are still disconnected
from the distant zones.  Integrating the
geodesic equations completely, we demonstrate the existence of
timelike and null geodesics connecting two asymptotic regions of the
wormhole, such that the tidal forces in the throat are reasonably
small. We also discuss bounds on the NUT charge which follow
from the Schwinger pair creation and ionization thresholds.

\end{abstract}
\pacs{04.20.Jb, 04.50.+h, 04.65.+e}
\maketitle
\setcounter{equation}{0}
\section{Introduction}
In this paper we would like to explore a novel aspect of non-vacuum
space-times endowed with a Newman-Unti-Tamburino (NUT) parameter.
Recall that the vacuum Taub solution was first presented as an
anisotropic cosmological solution \cite{Taub}, and later
rediscovered by Newman, Tamburino and Unti \cite{NUT} as a static
black hole (for more details see \cite{hawking}). This latter
solution has a unique event horizon of the Schwarzschild type, the
corresponding internal metric being just the original Taub
cosmology. Soon after Brill presented a corresponding pair for the
Einstein-Maxwell system \cite{brill}, its black hole face is
commonly called the Reissner-Nordstr\"om-NUT (RN-NUT) solution.
Depending on the relative values of parameters, it can have  two
horizons, one degenerate horizon, or no horizon. The cosmological
Brill solution corresponds to the region between two horizons in the
first case. Here we investigate the third case of no horizon, which
surprisingly was not discussed in the literature so far. It
corresponds to a Lorentzian wormhole connecting two asymptotically
locally flat spaces. Because it is free of a central singularity,
there is no place for a mass or charge, so that the Brill wormhole
is a realization of Wheeler's ``charge without charge'' and ``mass
without mass'' \cite{wheeler}.

An astonishing feature of this ``NUT wormhole'' is that it does not
demand an exotic matter  violating the null energy condition (NEC),
or the averaged NEC (ANEC), crucial for the existence of spherical
\cite{MT1,MT2,Visser} or non spherical \cite{teo} asymptotically
flat wormholes. Indeed, it is supported by a Maxwell field which is
not exotic. But the price for that is the presence of a metric
singularity on the polar axis known as the Misner string, which is
analogous to the Dirac string of the magnetic monopole in
electrodynamics. To make the string unobservable Misner suggested to
impose periodicity of the time coordinate \cite{misner1}, which
entails quantization of the energy of matter fields
\cite{Dowker,Perry} similar to Dirac charge quantization. This,
however, leads to violation of causality throughout the space-time,
so it can hardly be considered as a physically acceptable condition.
In the black hole case, another problem is that analytic
continuation cannot be consistently carried out through both
horizons, so that the resulting spacetime is geodesically
incomplete. In the wormhole case there are no horizons and analytic
continuation just reduces to the extension of the semiaxis or the
radial variable $r$ to the whole axis. So in the case of the Misner
interpretation the wormhole RN-NUT spacetime is geodesically
complete.

In an alternative interpretation of the Misner string suggested by
Bonnor \cite{bonnor} (see also \cite{Sut}), the time periodicity
condition is abandoned and the polar axis is treated as a kind of
topological defect with its proper matter source. We will assume
this viewpoint here. Since the polar axis is now an unremovable
singularity of spacetime, it was expected so far that geodesics will
stop here. Recently we have shown \cite{Clement:2015cxa} that it is
not true for the pure Taub-NUT metric. Here we extend this proof to
the Brill charged solution as well. Our proof is independent on
whether one assumes the Misner or Bonnor interpretation of the
space-time. In the wormhole case the space-time structure is
particularly simple, and since there is no singularity other than
the polar axis this means that the Brill spacetime is geodesically
complete.

Also, in the Bonnor interpretation there is a region around the
Misner string where closed timelike curves (CTCs) exist. Recently we
have shown that in the case of the Taub-NUT spacetime these CTCs are
{\em not geodesic} for a suitable choice of a parameter fixing the
position of the Misner string. Thus a freely falling observer will
not be confronted with causality violation. Here we extend the proof
to the Brill solution too. Since we deal with charged wormholes, it
is natural to consider also the worldlines of  charged particles. We
show that a circular worldline lying at the throat of a massless
magnetically charged wormhole can be causality violating for large
enough charge-to-mass ratio and a certain sign of the charge. At the
same time, we show that all the wordlines extending to large
distances from the throat are causal. Together with the proof of
geodesic completeness, this makes us believe that  NUT wormholes are
free from traditional objections against solutions with NUTs.

Passing to the analysis of the physical features of the solutions,
we first demonstrate the existence of timelike and null geodesics
connecting two asymptotic zones. Then we investigate the tidal
forces in the vicinity of the throat and show that these may be
reasonably small for a large NUT charge. More restrictive bounds on
this parameter are obtained by demanding that the electric field in
the throat is lower than the Schwinger pair creation or the matter
destruction characteristic fields. We conclude with some suggestions
for further theoretical work.

\setcounter{equation}{0}
\section{The Brill solution}
The solution of the Einstein-Maxwell system of equations found by
Brill in 1964 \cite{brill} soon after the discovery of Newman,
Tamburino and Unti \cite{NUT} reads
 \ba\lb{nworm}
ds^2 &=& - f(dt-2n(\cos\theta+C)\,d\varphi)^2 + f^{-1}dr^2 +
(r^2+n^2)(d\theta^2 + \sin^2\theta d\varphi^2)\,, \nn\\
A &=& \Phi(dt-2n(\cos\theta+C)\,d\varphi)\,,
 \ea
with
 \ba\lb{fem}
f &=& \frac{(r-m)^2+b^2}{r^2+n^2}\,, \quad \Phi = \frac{qr +
p(r^2-n^2)/2n}{r^2+n^2}\,, \\ && (b^2 = q^2+p^2-m^2-n^2)\,. \nn
 \ea
The metric depends only on the combination $e^2 = q^2 + p^2$ of the
electric ($q$) and magnetic ($p$) charges. For $e=0$
($b^2=-(m^2+n^2)$) the solution reduces to the Taub-NUT solution.
For $n=0$ ($b^2 = e^2-m^2$), it reduces to the dyonic
Reissner-Nordstr\"om solution. It is particularly simple in the
massless case  $m=0$ for the charges satisfying  $e^2=2n^2$
($b^2=n^2$), then the gravistatic potential $f(r)=1$, and the
solution looks like a ``pure'' NUT. Finally, for $q^2+p^2=m^2+n^2$
($b=0$), the solution is a special case of the
Israel-Wilson-Perj\`es (IWP) solution:
 \ba
ds^2 &=& -f\left(dt+\vec{\cal A}.d\vec{x}\right)^2 +
f^{-1}d\vec{x}^2\,, \quad A = \Phi\left(dt+\vec{\cal
A}.d\vec{x}\right)\,,\nn\\
f^{-1} &=& 1 + 2\sigma\cos\alpha + \sigma^2\,,  \quad
\nabla\wedge\vec{\cal A} = 2\sin\alpha\nabla\sigma\,,\nn\\
\Phi &=& \frac{f^{-1}} {2\sin\alpha}\,[\sin\beta +
2\sigma\sin(\alpha+\beta) + \sigma^2\sin(2\alpha+\beta)]\,,
 \ea
where $\sigma(\vec{x})$ a harmonic function.

The value of the parameter $C$ introduced above can be modified by a
``large'' coordinate transformation
 \be\lb{tgauge}
t \to t + 2n(C-C')\varphi\,.
 \ee
Such a transformation will generically lead to an intrinsically
different spacetime. Thus, (\ref{nworm}) actually defines a
one-parameter family of spacetimes, to which we will refer
collectively as ``the Brill spacetime''. We will keep this parameter
$C$ free until section 6.

In the case of a non-vanishing NUT charge, $n\neq0$, the Brill
solution (\ref{nworm}) is not singular at $r=0$, as can be checked
by computing the quadratic curvature invariants,
 \ba
& R^{\mu\nu}R_{\mu\nu} & = \frac{4e^4}{(r^2+n^2)^4}\;, \lb{R2}\\
& R^{\mu\nu\rho\sigma}R_{\mu\nu\rho\sigma} & =
\frac{8e^4}{(r^2+n^2)^4} +
\frac{48}{(r^2+n^2)^6}\left\{(m^2-n^2)\,[r^6-15n^2r^4+15n^4r^2-n^6]
\right. \nn\\  && \left. -2mr\,[(e^2-6n^2)r^4 - (10e^2-20n^2)n^2r^2
+ (5e^2-6n^2)n^4] \right. \nn\\ && \left. + e^2[(e^2-10n^2)r^4 -
2(3e^2-10n^2)n^2r^2 + (e^2-2n^2)n^4] \right\}\,. \lb{K}
 \ea
The Kretschmann scalar (\ref{K}) (where the first and second terms
are respectively the Ricci square and Weyl square contributions),
reduces to that of the Reissner-Nordstr\"om solution \cite{HSS} for
$n^2=0$, and to that of the Schwarzschild-NUT solution \cite{vish}
for $e^2 = 0$. In view of its non-singularity, the Brill solution
can be considered as a regularization of the Reissner-Nordstr\"om
solution. For $b^2<0$ it has, just as the RN solution, two horizons.
For $b^2 = 0$, it has, just as the extreme RN solution, a double
horizon. However for $b^2 > 0$, contrary to the RN solution, it is
not singular, but has the (Lorentzian) wormhole topology, the
coordinate $r$ varying in the whole real axis, with two asymptotic
regions $r=\pm\infty$. As $r>0$ decreases, 2-spheres $r=$ constant
shrink until a minimal sphere of area $4\pi n^2$ (the wormhole neck)
for $r=0$, and then expand as $r<0$ continues to decrease. Because
of the absence of a point singularity, there is no source for the
various (gravi-)electric and (gravi-)magnetic fluxes, and thus this
solution realizes the Wheeler program \cite{wheeler} of mass without
mass, charge without charge, etc. (for a more recent discussion see
\cite{Guen}. Note also that (as in the Taub-NUT case $e=0$), the
mass parameter $m$ is not positive definite, as the reflection
$r\to-r$ changes its sign and that of the electric charge $q$.

The price to pay for this regularization of the RN point singularity
is the introduction of the Misner string singularity. The ``Misner
string'' (which coincides with the Dirac string for $p\neq0$)
consists of two disconnected infinite components, the North string
piercing the North poles $\theta=0$ of all the spheres $r =$
constant ($r\in R$) and its counterpart the South string for
$\theta=\pi$, except in the special gauges $C=\mp1$, where only one
component (South or North) is singular. The Misner string
singularity can be transformed away altogether if the time
coordinate $t$ is periodically identified with period $8n\pi$.
However there is also a price to pay for this. First, there are
closed timelike curves (CTC) everywhere in all the Brill spacetimes.
Second, in the case $b^2\le0$, the Kruskal continuation cannot be
consistently carried out with this periodical identification, so
that the black-hole Brill spacetimes are not geodesically complete
\cite{hawking,kagra}, i.e. the Misner string singularity has been
traded for a singular horizon.

So let us keep the usual real time axis, thereby retaining also the
Misner string. In this case it has been argued \cite{kagra} (in the
case of the Taub-NUT black hole spacetime, but exactly the same
argument could be formulated in the general Brill case) that
geodesics must terminate at the Misner string singularity, just
because this is a metric singularity. Actually, geodesic motion does
not depend on the metric itself, but only on the connections, so a
metric singularity does not necessarily imply geodesic termination.
Geodesics do indeed terminate at a conical singularity (the analogy
made in \cite{kagra}), but not at the metric singularity $r=0$ of
the plane with metric $dl^2 = dr^2+r^2d\theta^2$! The investigation
of geodesic motion in the metric (\ref{nworm}), carried out in the
next section, will show 1) that all geodesics which hit the Misner
string cross it smoothy, so that all the Brill spacetimes with
$n^2\neq0$ are geodesically complete; 2) that if furthermore $b^2 >
0$ the resulting Lorentzian wormholes are traversable.

\setcounter{equation}{0}
\section{Geodesics: Angular and temporal motion}

The partially integrated geodesic equations for the metric
(\ref{nworm}) are \cite{ZS}
 \ba
&& f(r)(\dot{t} - 2n(\cos\theta+C)\,\dot\varphi) = E\,, \lb{t}\\
&& \dot\varphi = \frac{L_z(\theta)}{(r^2+n^2)\sin^2\theta}\,, \lb{varphi}\\
&& [(r^2+n^2)\dot\theta]\,\dot{} = (r^2+n^2)\sin\theta\cos\theta
\,\dot\varphi^2 - 2nE\sin\theta\,\dot\varphi\,, \lb{theta0}\\
&& f(r)^{-1}(\dot{r}^2-E^2) + (r^2+n^2)(\dot\theta^2 +
\sin^2\theta\,\dot\varphi^2) = \varepsilon\,, \lb{rad}
 \ea
where $\dot{}= d/d\tau$, with $\tau$ an affine parameter,
 \be
L_z(\theta) = J_z - 2nE\cos\theta\,,
 \ee
and $E$ and $J_z$ are the constants of the motion associated with
the cyclic variables $t$ and $\varphi$ (see Appendix A). In the
following we assume without loss of generality that for timelike and
null geodesics $E>0$, thereby defining the orientation of time, and
$n>0$, the sign of $J_z$ remaining arbitrary. Eq. (\ref{rad}) is
$ds^2 = \varepsilon d\tau^2$, with $\varepsilon=-1$, $0$ or $1$ for
timelike, null or spacelike geodesics, respectively.

The analysis of geodesic motion in the Brill metric parallels that
of a charged particle in the field of a magnetic monopole \cite{ZS}.
Similarly to the case of the Taub-NUT metric, the metric
(\ref{nworm}) admits four Killing vectors, one ($K_0$) generating
time translations, and three ($K_i$) generating the rotation group
$O(3)$. Due to this spherical symmetry, the angular components of
the geodesic equations can be first integrated by
 \be\lb{J}
\vec{L} + \vec{S} = \vec{J}\,,
 \ee
where $\vec{J}$ is a constant vector, the total angular momentum,
which is the sum of an orbital angular momentum $\vec{L}$ and a
``spin'' angular momentum $\vec{S}$. The components of $\vec{L}$ and
$\vec{S}$ are given in \cite{ZS}, where it is shown that
 \be
\vec{L} = (r^2+n^2)\,\r\wedge\dot{\r}\,, \qquad \vec{S} =
2nE\hat{r}\,
 \ee
where
 \be
\r = (\sin\theta\cos\varphi,\,\sin\theta\sin\varphi,\,\cos\theta)
 \ee
is a unit vector normal to the two-sphere.

It follows from the orthogonality of $\vec{L}$ and $\vec{S}$ that
 \be\lb{para}
\vec{J}\cdot\hat{r}=2nE\,.
 \ee
In the magnetic monopole case, such a first integral means that the
the trajectory of the charged particle lies on the surface of a cone
with axis $\vec{J}$ originating from the magnetic monopole source
$r=0$. However the Taub-NUT or Brill gravitational field has no
source, i.e. no apex for the ``cone''. Actually, the content of
(\ref{para}) is that the geodesic intersects all the two-spheres of
radius $r$ on the same small circle, or parallel, $\C$ with polar
axis $\vec{J}$. Squaring (\ref{J}) leads to
 \be\lb{J2LE}
\vec{J}^2 = \vec{L}^2 + 4n^2E^2\,,
 \ee
which can be rewritten as
 \be\lb{l2}
(r^2+n^2)^2[\dot\theta^2 + \sin^2\theta\,\dot\varphi^2] = l^2\,,
 \ee
with $l^2 = J^2-4n^2E^2$ ($J^2 = \vec{J}^2$). Insertion into
(\ref{rad}) leads to the effective radial equation
 \be\lb{rad1}
\dot{r}^2 + f(r)\left[\frac{l^2}{r^2+n^2}-\varepsilon\right] =
E^2\,,
 \ee
which is identical to the equation for radial motion in the
equatorial plane for the metric (\ref{nworm}) without the term
$-2n\cos\theta d\varphi$, i.e. for $b^2>0$ motion in a static
spherically symmetric wormhole geometry of the Ellis type \cite{kb}
(the $m=0$ symmetrical Ellis wormhole \cite{ellis} for $f(r) = 1$).

\subsection{Angular motion}
Knowing (in principle) the solution to (\ref{rad1}), one can insert
it in (\ref{l2}), where $\dot\varphi$ can be eliminated by using
(\ref{varphi}), to yield a first-order differential equation for
$\theta(\tau)$. Then the solution can be used to obtain
$\varphi(\tau)$ and $t(\tau)$ by integrating (\ref{varphi}) and
(\ref{t}). The Misner string singularities $\theta=0$, $\pi$ do not
seem to play any role in the complete integration thus carried out.
To check this, we follow \cite{kagra} and replace the affine
parameter $\tau$ by the new variable $\lambda$ defined by
\begin{equation}
d\tau =\left( r^{2}+n^{2}\right) d\lambda\,,   \label{mint}
\end{equation}
which increases monotonously with $\tau$. Putting
\begin{equation}
\xi =\cos \theta\,,
\end{equation}
the differential equation for $\theta(\lambda)$ reads
\begin{equation}\lb{theta}
\left(\frac{d\xi}{d\lambda}\right)^{2} = -J^{2}\,\xi^2 +
4nEJ_{z}\,\xi + (l^{2}-J_{z}^{2})\,.
\end{equation}

Assuming $J^2\neq0$ ($J^2=0$ implies from (\ref{J2LE}) $E=0$ and
$l=0$), Eq. (\ref{theta}) is solved (up to an additive constant to
$\lambda$) by \cite{kagra}
 \be\lb{theta1}
\cos\theta = J^{-2}\left[2nEJ_z + lJ_{\bot}\cos(J\lambda)\right]\,,
 \ee
where
 \be\lb{Jbot}
J_{\bot}^2 = J^2 - J_{z}^2\,.
 \ee
Because the constants of the motion occurring in (\ref{theta1}) are
related by the two Pythagorean decompositions of $\vec{J}^2$
(\ref{J2LE}) and (\ref{Jbot}), it will be useful to trade them for
the two angles $\eta$ and $\psi$ defined by
 \be\lb{angles}
2nE = J\cos\eta\,, \quad l = J\sin\eta\,, \quad J_z = J\cos\psi\,,
\quad J_{\bot} = J\sin\psi\,,
 \ee
with $0\le\eta\le\pi/2$ and $0\le\psi\le\pi$. Then (\ref{theta1})
reads
 \be\lb{theta2}
\cos\theta = \cos\psi\cos\eta + \sin\psi\sin\eta\cos(J\lambda)\,.
 \ee

Eq. (\ref{theta}) has two turning points $\theta_{\pm}$ such that
 \be\lb{turn}
\cos\theta_{\pm} = J^{-2}\left(2nEJ_z \pm lJ_{\bot}\right) =
\cos(\psi\mp\eta)\,.
 \ee
These can be rewritten as
 \be\lb{thetapm}
\theta_+ = \left\vert \begin{array}{lcc} \psi-\eta & {\rm if} &
\psi\ge\eta \\ -\psi+\eta & {\rm if} & \psi\le\eta
\end{array}\right. \,, \quad \theta_- = \left\vert \begin{array}{ccc}
\psi+\eta & {\rm if} & \psi\le\pi-\eta \\ -\psi-\eta+2\pi & {\rm if}
& \psi\ge\pi-\eta \end{array}\right. \,.
 \ee
It follows that the trajectory crosses periodically the Misner
string, $\cos\theta_{\pm} = \pm1$ only if
 \be\lb{cross}
\psi=\eta\;\;(J_z = 2nE) \quad {\rm or} \quad \psi=\pi-\eta\;\;(J_z
= -2nE)\,.
 \ee
The only geodesics which can cross both components of the Misner
string are those with $\eta=\pi/2$ ($E=J_z=0$), leading to
$\dot{t}=\dot{\varphi}=0\,$; according to (\ref{rad1}), in the
stationary sector ($f(r)>0$) these can only be spacelike geodesics.
The trajectory can also stay on the Misner string component
$\theta=0$ or $\pi$ if (\ref{cross})) is satisfied with $\eta = 0$
($2nE=J$).

The differential equation (\ref{varphi}) for $\varphi$ can be
rewritten as
 \be\lb{varphi1}
\frac{d\varphi}{d\lambda}
=\frac12\left[\frac{J_z-2nE}{1-\cos\theta(\lambda)} +
\frac{J_z+2nE}{1+\cos\theta(\lambda)}\right]\,,
 \ee
with $\cos\theta(\lambda)$ given by (\ref{theta1}). This is solved
by \cite{kagra}
 \ba\lb{solvarphi}
\varphi - \varphi_0 &=&
\arctan\left[\frac{\cos\psi-\cos\eta}{1-\cos(\psi-\eta)}
\tan\,\frac{J\lambda}2\right] \nn\\
&+& \arctan\left[\frac{\cos\psi+\cos\eta}{1+\cos(\psi-\eta)}
\tan\,\frac{J\lambda}2\right]\,,
 \ea
with $\varphi_0$ an integration constant. This can be simplified to
 \be\lb{solvarphi1}
\varphi - \varphi_0 = \arctan\left[\frac{2\sin\eta\tan(J\lambda/2)}{
\sin(\eta-\psi) - \sin(\eta+\psi)\tan^2(J\lambda/2)}\right]\,.
 \ee
For trajectories crossing the North Misner string, with $\psi=\eta$
($J_z = 2nE$), this reduces to
 \be\lb{solvarphi2}
\varphi - \varphi_1 =
\arctan\left(\cos\eta\,\tan\left(\frac{J\lambda}2\right)\right)\,,
 \ee
with $\varphi_1 = \varphi_0 - {\rm{sgn}}(\tan(J\lambda/2))\pi/2$. A
similar formula applies in the case of the South Misner string, with
$\eta$ replaced by $\pi-\eta$ and $J\lambda$ replaced by
$J\lambda-\pi$ (note that according to (\ref{theta1}) the North
Misner string is crossed for $\lambda=2k\pi/J$, while the South
Misner string is crossed for $\lambda=(2k+1)\pi/J$, $k$ integer). In
the case e.g. of the North Misner string, this gives on account of
(\ref{theta1}),
 \be\lb{solvarphi3}
\cos(\varphi - \varphi_1) =
\frac{J_z}{J_{\bot}}\tan\left(\frac{\theta}2\right)\,,
 \ee
consistent with (\ref{para}) (the choice $\varphi_1=0$ in
(\ref{solvarphi3}) corresponds to the choice $\vec{J} =
(J_{\bot},0,J_z)$ in (\ref{para})). Clearly the Misner string is
completely transparent to the geodesic motion!

When the parameter $\lambda$ varies over a period, e.g. $\lambda \in
[-\pi/J,\,\pi/J]$, the argument of the first or second $\arctan$ in
(\ref{solvarphi}) varies from $-\infty$ to $+\infty$ for
$J_z\mp2nE>0$, and from $+\infty$ to $-\infty$ for $J_z\mp2nE<0$. It
is identically zero for $J_z\mp2nE=0$. Accordingly, the variation of
$\varphi$ over a period is
 \be\lb{Dvarphi}
\Delta\varphi = \pi \left[{\rm sgn}(J_z-2nE) + {\rm sgn}(J_z+2nE)
\right]\,.
 \ee
This means that for $J_z^2>4n^2E^2$ ($|\Delta\varphi|=2\pi$) the
parallel $\C$ circles the North-South polar axis, i.e. the Misner
string. For $J_z^2<4n^2E^2$, ($|\Delta\varphi|=0$) $\C$ does not
circle the Misner string. And for $J_z=\pm2nE$
($|\Delta\varphi|=\pi$), $\C$ goes through the North or South pole,
as discussed above.

The two turning points (\ref{thetapm}) coincide if either $\psi =
0$, or $\eta = 0$:

a) $\psi = 0$ or $\pi$. The constant vector $\vec{J}$ is aligned
with the $z$ axis, on which the parallel $\C$ is centered:
 \be\lb{framepsi0}
\theta = \eta\; {\rm or}\; \pi-\eta\,, \quad \varphi-\varphi_0 = \pm
J\lambda\,.
 \ee
As discussed in \cite{ZS}, the solution (\ref{theta1}),
(\ref{solvarphi1}) can always be rotated to this case, the Misner
string being then rotated away from the $z$ axis.

b) $\eta = 0$. The orbital angular momentum vanishes, $l=0$, and the
motion is  purely radial,
 \be
\theta = \psi\,, \quad \varphi-\varphi_0 = 0\,.
 \ee
This will be further discussed in section 4.

\subsection{Temporal motion}
After transforming to the geodesic variable $\tau$ by (\ref{mint}),
the solution to (\ref{t}) can be written as the sum
 \be
t(\lambda) = t_r(\lambda) + t_{\theta}(\lambda)\,,
 \ee
where the radial and angular contributions to $t(\lambda)$ solve the
equations
 \ba
\frac{dt_r}{d\lambda} &=& E\frac{r^2+n^2}{f(r)}\,, \lb{tr}\\
\frac{dt_{\theta}}{d\lambda} &=& \frac{2n(\cos\theta+C)(J_z -
2nE\cos\theta)}{\sin^2\theta}\lb{ta}\,.
 \ea
The solution to equation (\ref{tr}) depends on the solution of the
equation for radial motion (\ref{rad1}). We consider here equation
(\ref{ta}). This can be rewritten as
 \be\lb{ta1}
\frac{dt_{\theta}}{d\lambda} = 4n^2E +
n\left[\frac{(C+1)(J_z-2nE)}{1-\cos\theta(\lambda)} +
\frac{(C-1)(J_z+2nE)}{1+\cos\theta(\lambda)}\right]\,,
 \ee
with $\cos\theta(\lambda)$ given by (\ref{theta1}). The explicit
solution to equation (\ref{ta1}) is \cite{kagra}
 \ba\lb{solta1}
t_{\theta}(\lambda) &=& \ol{t}_{\theta}(\lambda) = 4n^2E\lambda +
2n(C+1) \arctan\left[\frac{\cos\psi-\cos\eta}{1-\cos(\psi-\eta)}
\tan\,\frac{J\lambda}2\right] \nn\\
&+& 2n(C-1)\arctan\left[\frac{\cos\psi+\cos\eta}{1+\cos(\psi-\eta)}
\tan\,\frac{J\lambda}2\right]\,,
 \ea
in the interval $-\pi/J < \lambda < \pi/J$. The resulting variation
of $t_{\theta}$ over a period $2\pi/J$ of $\lambda$ is
 \be\lb{Dttheta}
\Delta t_{\theta} = 2\pi n\left[\frac{4nE}J  + (C+1){\rm
sgn}(J_z-2nE) + (C-1){\rm sgn}(J_z+2nE) \right]\,,
 \ee
so that the solution for generic values of $\lambda$ is
 \be
t_{\theta}(\lambda) = \ol{t}_{\theta}(\lambda) + \left[\Delta
t_{\theta} - \frac{8\pi n^2E}J\right]{\rm
E}\left(\frac{J\lambda}{2\pi}+\frac12\right).
 \ee
For parallels $\cal C$ which do not circle the Misner string
($J_z^2<4n^2E^2$), $\Delta t_{\theta}$ is negative and independent
of the value of $C$, $\Delta t_{\theta} = - 4\pi n(1-\cos\eta)$. In
the limit when such parallels contract to a point, the motion
becomes purely radial ($2nE = J$) and the angular contribution
$t_{\theta}$ vanishes.

Let us recall here that, while the solution for the spatial
components of the geodesic (the orbit) can be rotated to a frame in
which the $z$ axis points along the total angular momentum vector
$\vec J$, this is not the case for the time coordinate $t(\lambda)$.
The correct equation for the time evolution in this rotated frame is
given (for the choice $C=0$) in \cite{ZS}, Eq. (24a).

\setcounter{equation}{0}

\section{Geodesics: The complete orbits}

\subsection{The radial potential}
The radial equation (\ref{rad1}) can be written in the familiar form
 \be\lb{rad2}
\left(\frac{dr}{d\tau}\right)^2 + U(r) = E^2+\varepsilon\,,
 \ee
with the effective potential
 \be\lb{Upot}
U(r) = \varepsilon\,\frac{2mr+n^2-\alpha^2}{\rho^2 } +
\frac{l^2[(r-m)^2 + b^2]}{\rho^4 }
 \ee
with $\alpha^2 = b^2+m^2 = e^2-n^2$. Its derivative generically can
presented as $U'=P_4(r)/\rho^6$, where $P_4(r)$ is a polynomial of
fourth order in $r$, which, depending on the parameter values, may
have four real roots or two real roots. In the cases $m=0$ or
$\varepsilon=0$, $P_4(r)$ degenerates to a polynomial of third order
with three or one real roots.
\begin{figure}[tb]
\begin{center}
\begin{minipage}[t]{0.48\linewidth}
\hbox to\linewidth{\hss%
  \includegraphics[width=0.95\linewidth,height=0.7\linewidth]{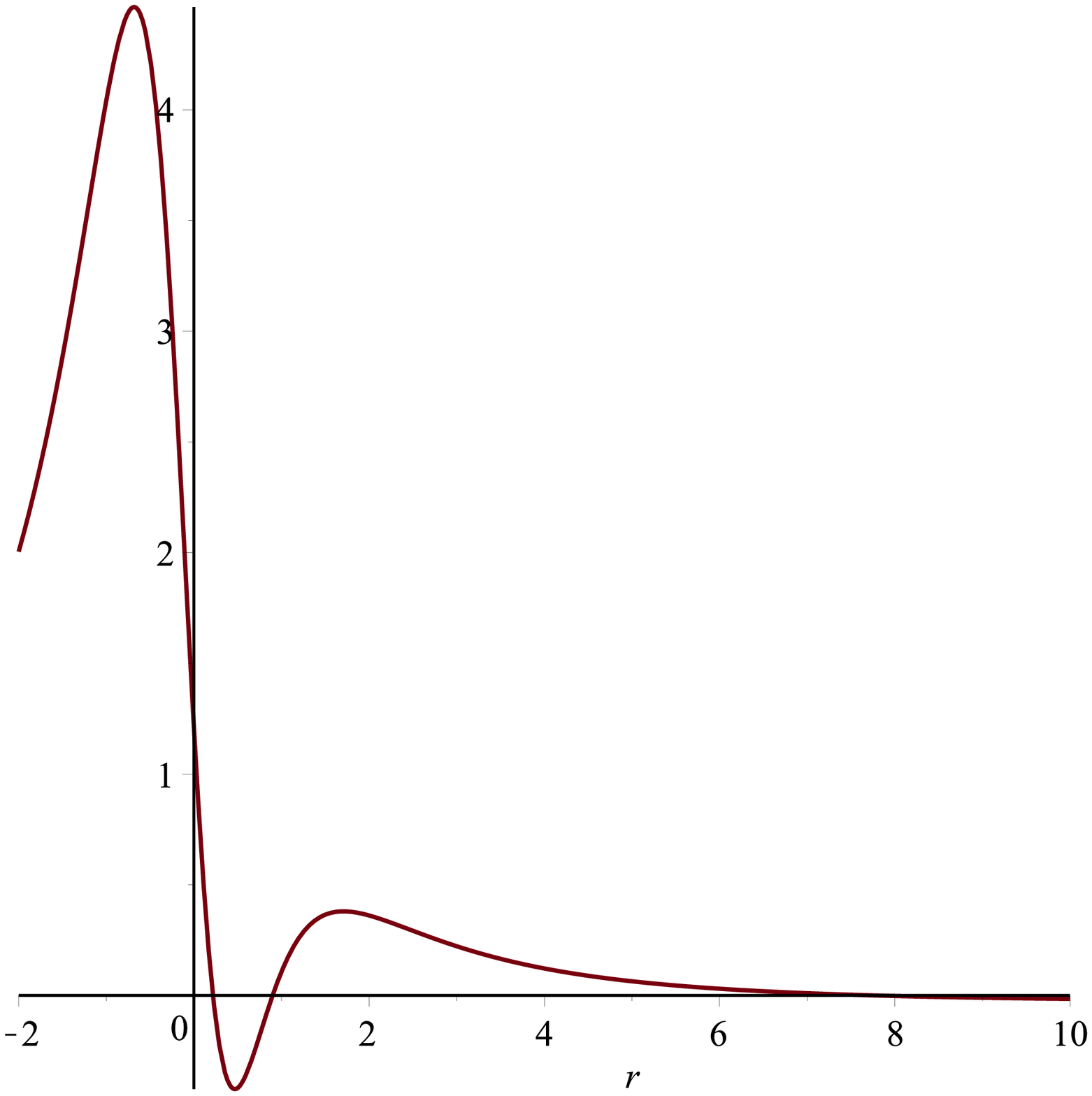}
\hss} \caption{\small  Radial potential corresponding to $n=1,\,
m=.4,\, l=2.8,\, b=.3$ with two maxima and two minima. The rightest
one at $r=16.23395$, $U=-0.2383e-1$, not seen in this scale, is
shown separately in Fig. 2.} \label{F1}
\end{minipage}
\hfill
\begin{minipage}[t]{0.48\linewidth}
\hbox to\linewidth{\hss%
  \includegraphics[width=0.95\linewidth,height=0.7\linewidth]{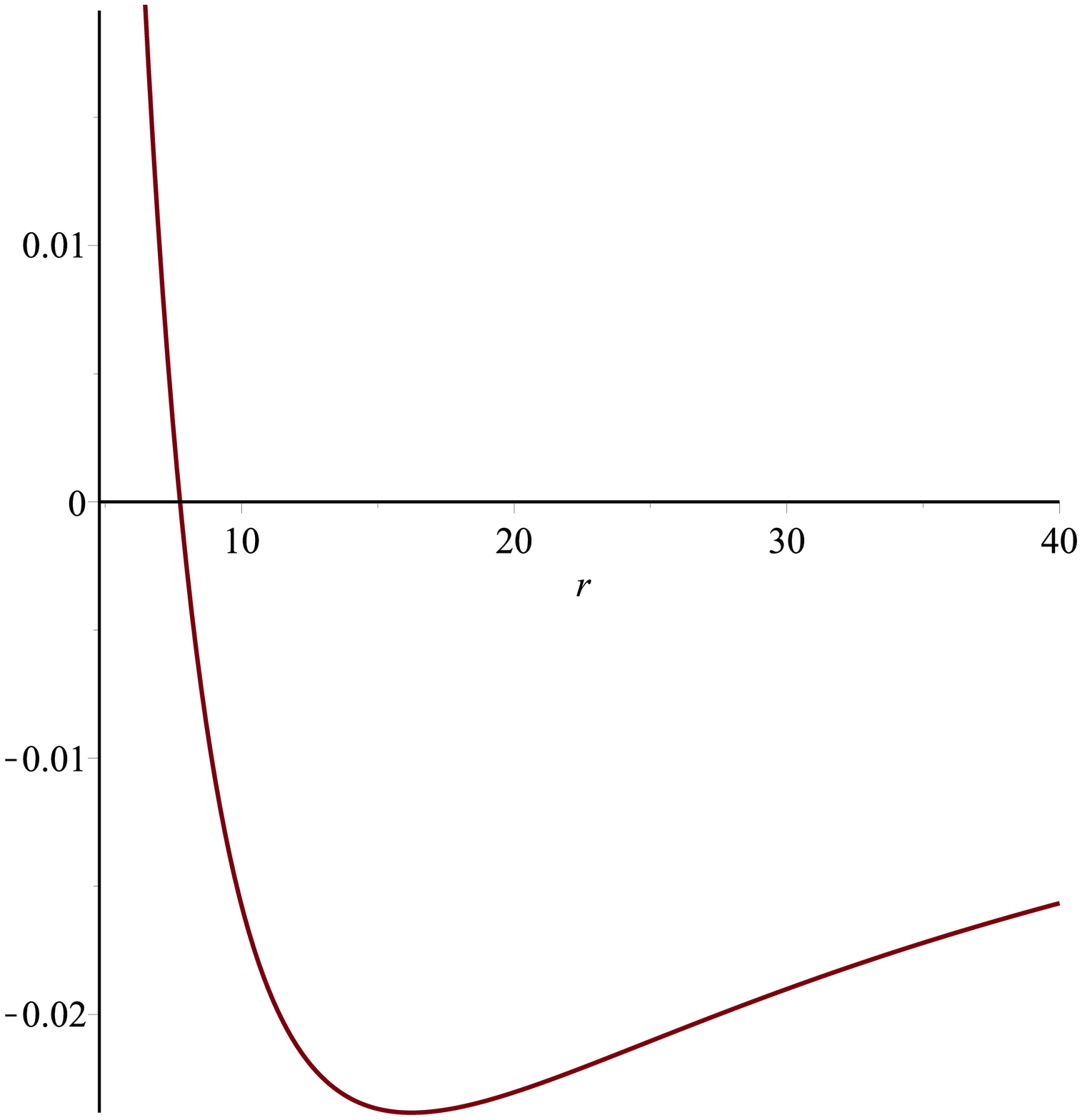}
\hss} \caption{\small  The Newtonian potential well region of the
potential Fig. 1  with higher resolution.} \label{F2}
\end{minipage}
\end{center}
\end{figure}

\begin{figure}[tb]
\begin{center}
\begin{minipage}[t]{0.48\linewidth}
\hbox to\linewidth{\hss%
  \includegraphics[width=0.95\linewidth,height=0.7\linewidth]{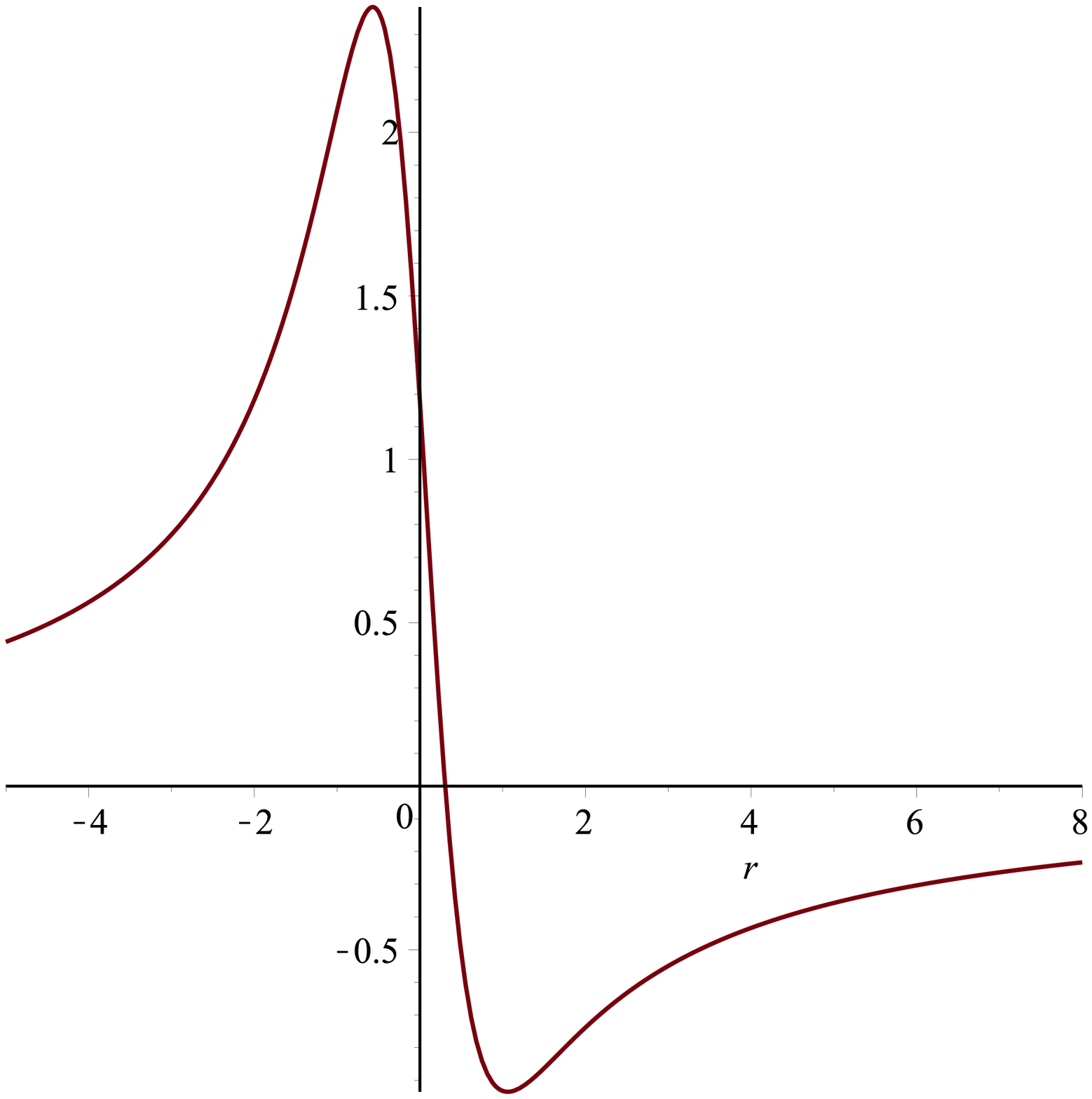}
\hss} \caption{\small The potential with two extrema for non-zero
orbital momentum ($n=m=l=1,\, b=.3$).  } \label{F3}
\end{minipage}
\hfill
\begin{minipage}[t]{0.48\linewidth}
\hbox to\linewidth{\hss%
  \includegraphics[width=0.95\linewidth,height=0.7\linewidth]{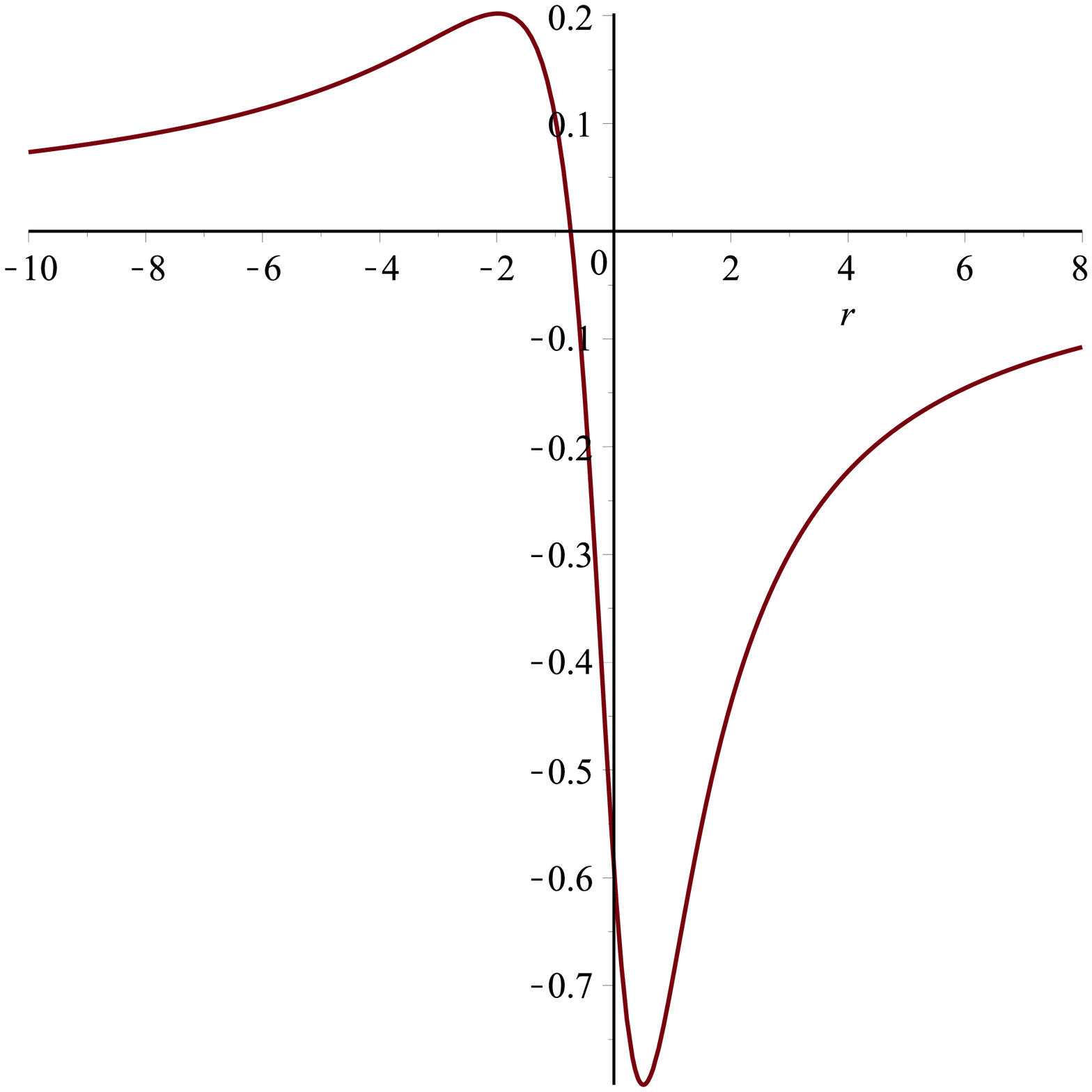}
\hss} \caption{\small  The potential for zero orbital momentum $l=0$
and $n=1\,, m=.4\,, b=.5$. For $l=0$ the extrema can be found
analytically by solving a quadratic equation.} \label{F4}
\end{minipage}
\end{center}
\end{figure}

\begin{figure}[tb]
\begin{center}
\begin{minipage}[t]{0.48\linewidth}
\hbox to\linewidth{\hss%
  \includegraphics[width=0.95\linewidth,height=0.7\linewidth]{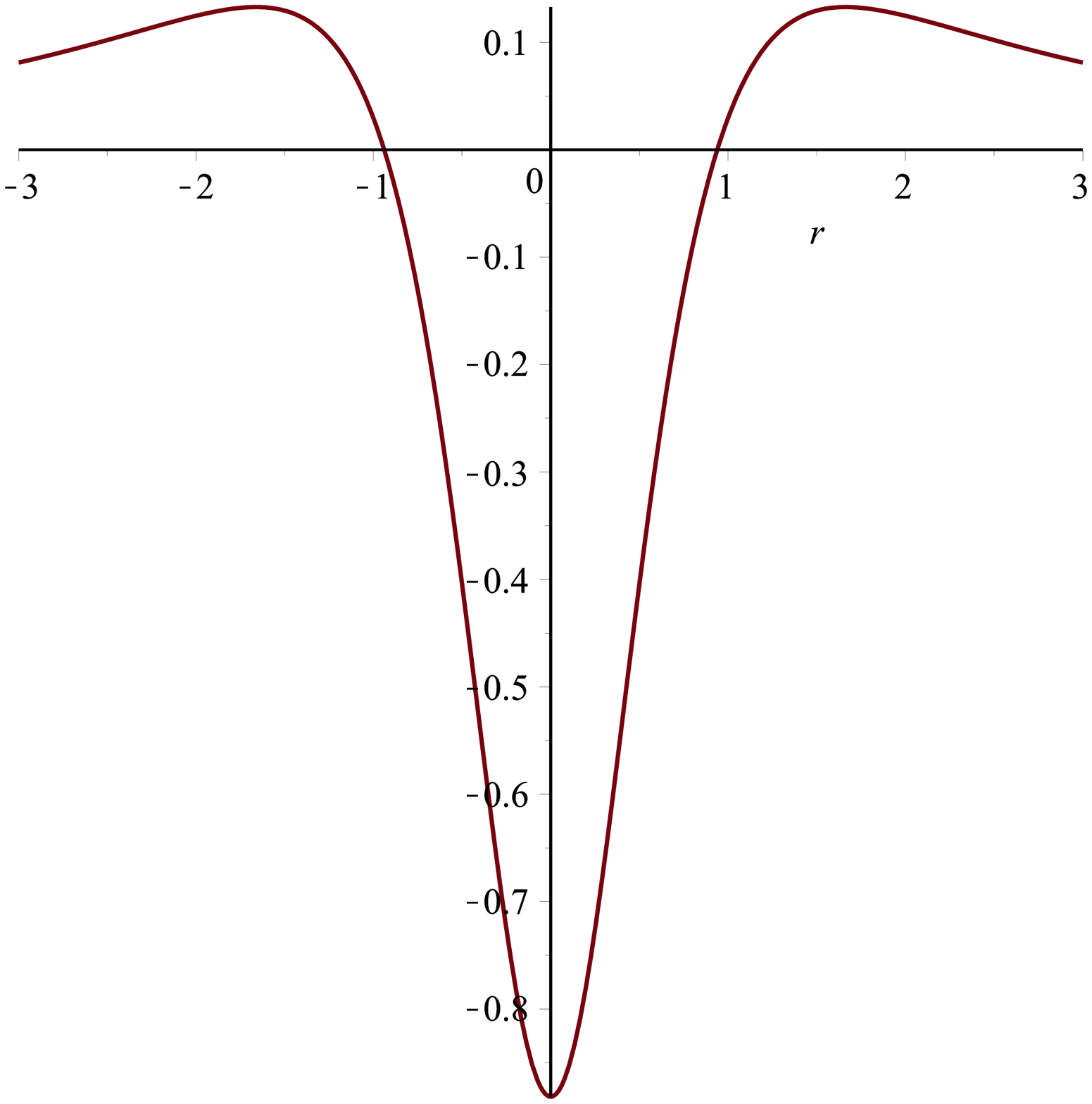}
\hss} \caption{\small The generic potential for zero mass $m=0$ with
three extrema  and  $n=1\,, l=1\,, b=.2$} \label{F5}
\end{minipage}
\hfill
\begin{minipage}[t]{0.48\linewidth}
\hbox to\linewidth{\hss%
  \includegraphics[width=0.95\linewidth,height=0.7\linewidth]{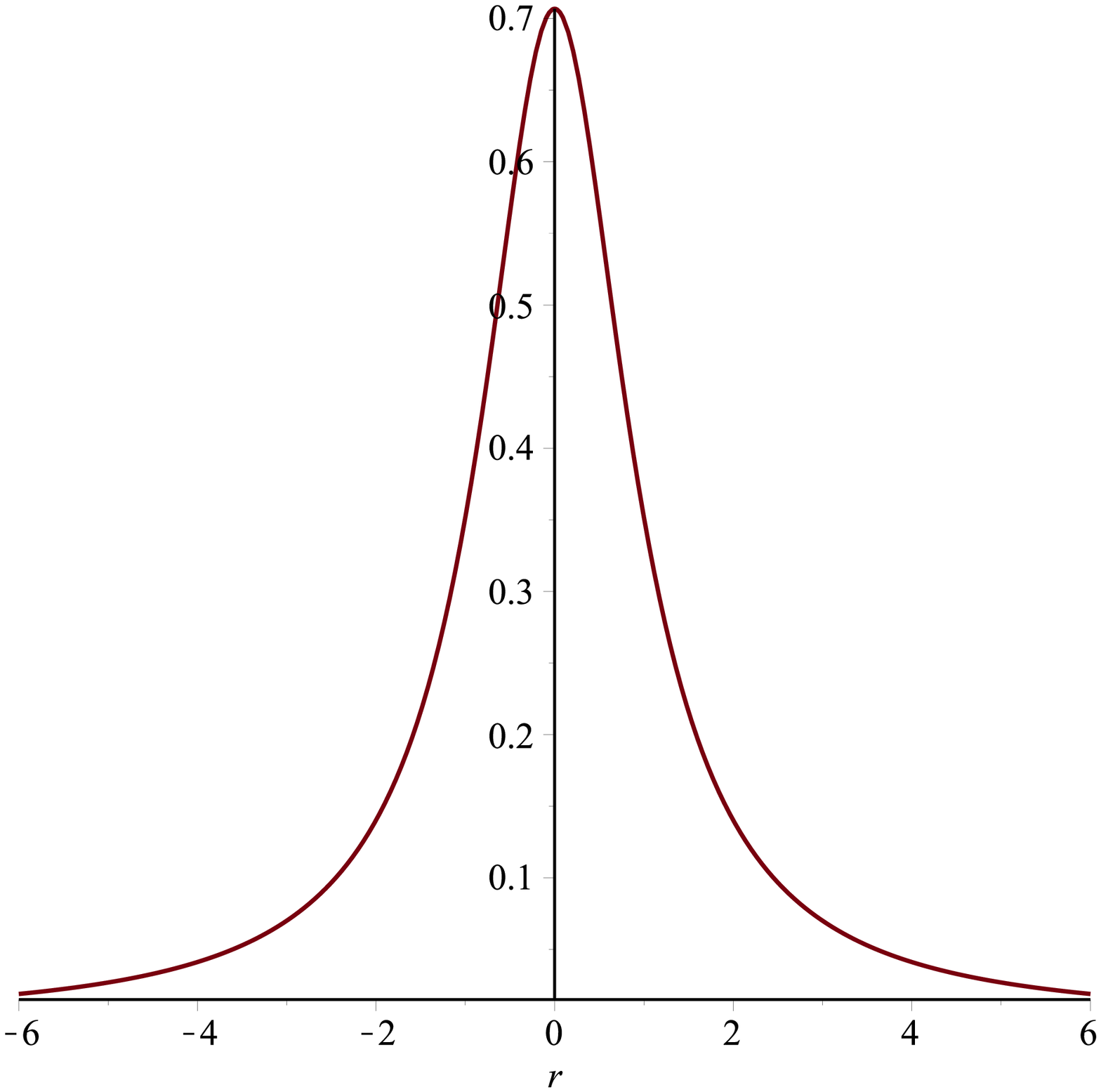}
\hss} \caption{\small Massless wormhole potential with one maximum
($m=0,\,l=0.1,\,n=1,\, b=1.3$). } \label{F6}
\end{minipage}
\end{center}
\end{figure}

\begin{figure}[tb]
\begin{center}
\begin{minipage}[t]{0.48\linewidth}
\hbox to\linewidth{\hss%
  \includegraphics[width=0.95\linewidth,height=0.7\linewidth]{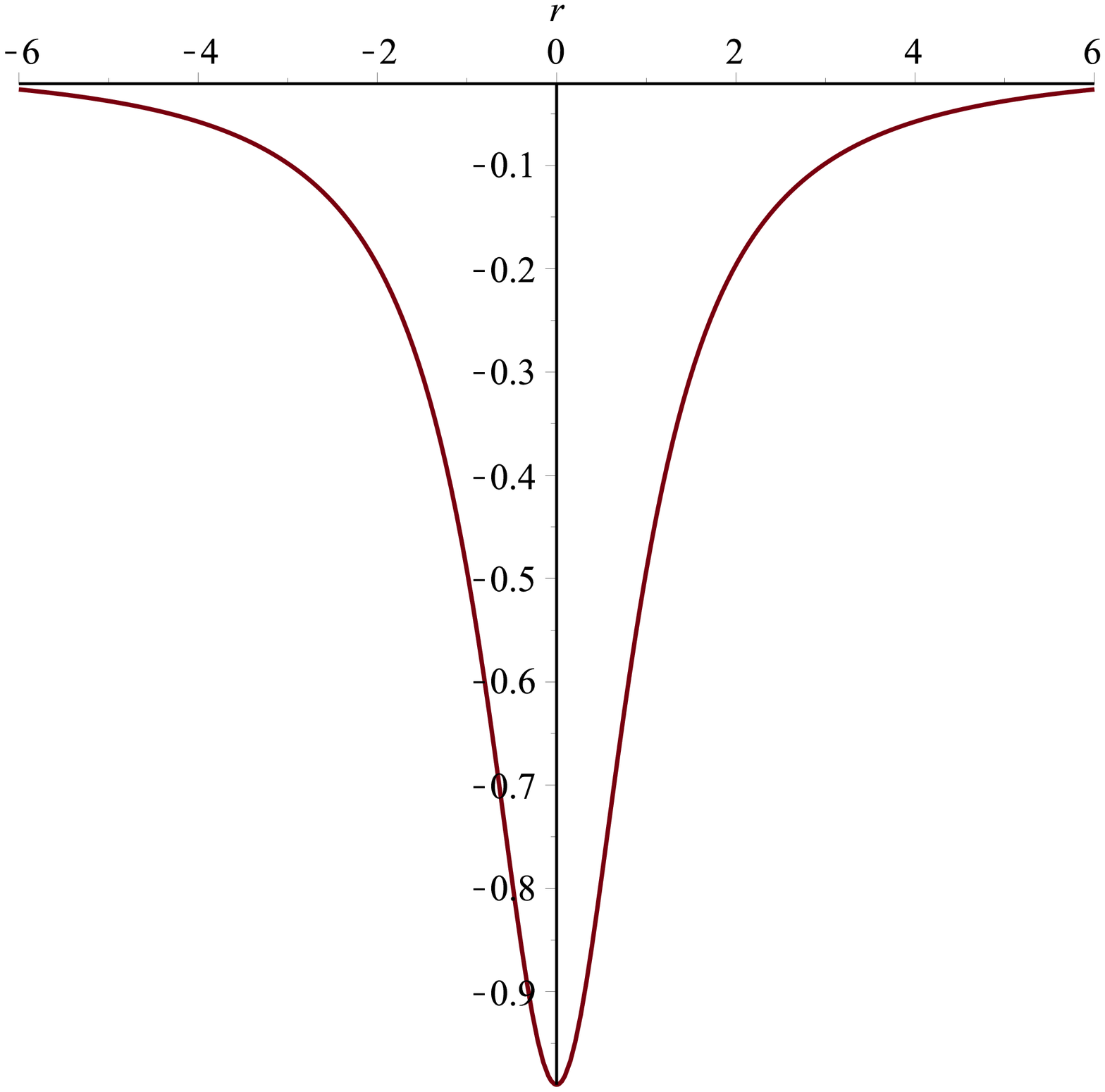}
\hss} \caption{\small Massless wormhole potential with one minimum
($m=0,\,l=.1,\,n=1,\, b=0.1$.} \label{F7}
\end{minipage}
\hfill
\begin{minipage}[t]{0.48\linewidth}
\hbox to\linewidth{\hss%
  \includegraphics[width=0.95\linewidth,height=0.7\linewidth]{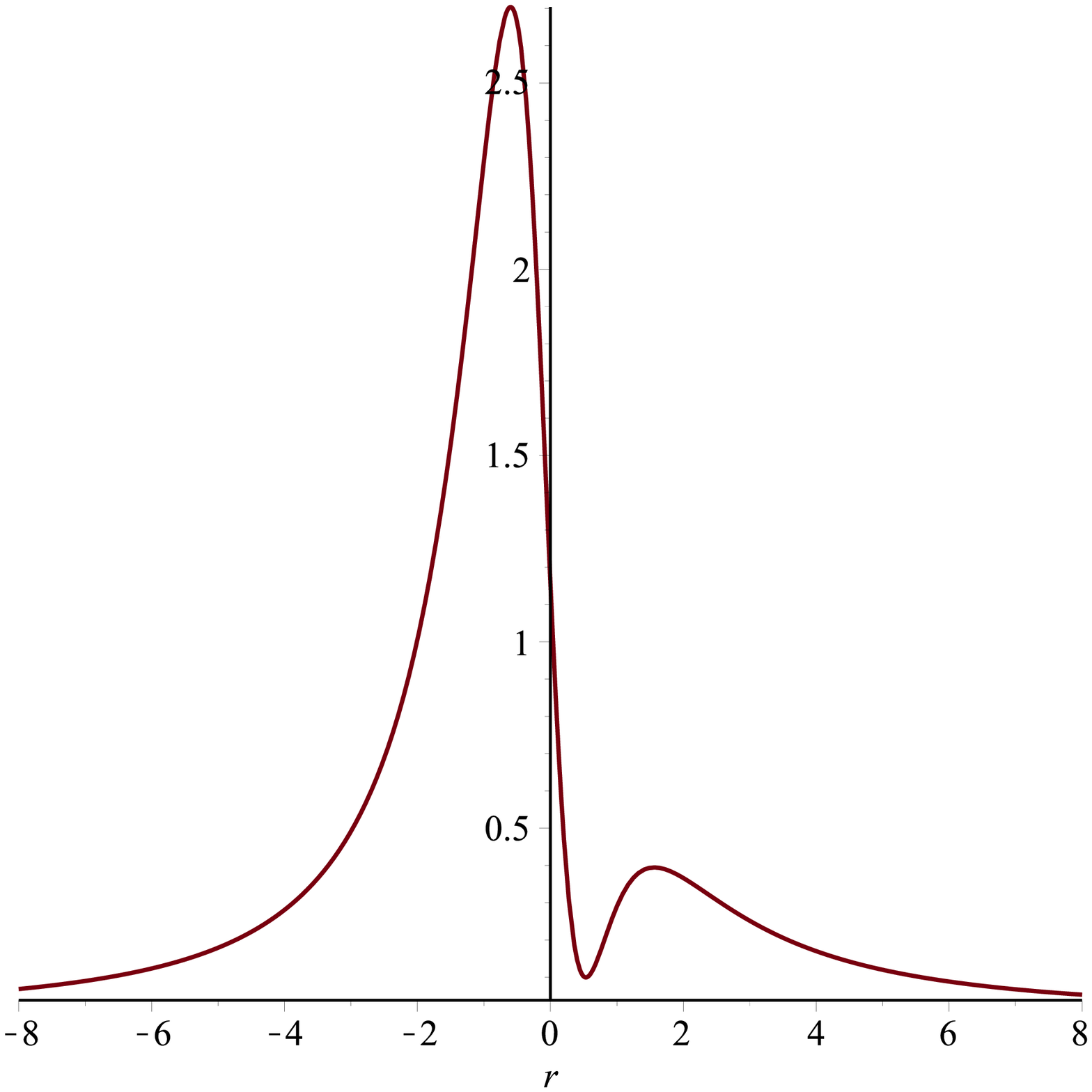}
\hss} \caption{\small  Generic potential for null geodesics with
three extrema ($n=1,\, m=1/2,\, l=2,\, b=.2$).   } \label{F8}
\end{minipage}
\end{center}
\end{figure}

\begin{figure}[tb]
\begin{center}
\begin{minipage}[t]{0.48\linewidth}
\hbox to\linewidth{\hss%
  \includegraphics[width=0.95\linewidth,height=0.7\linewidth]{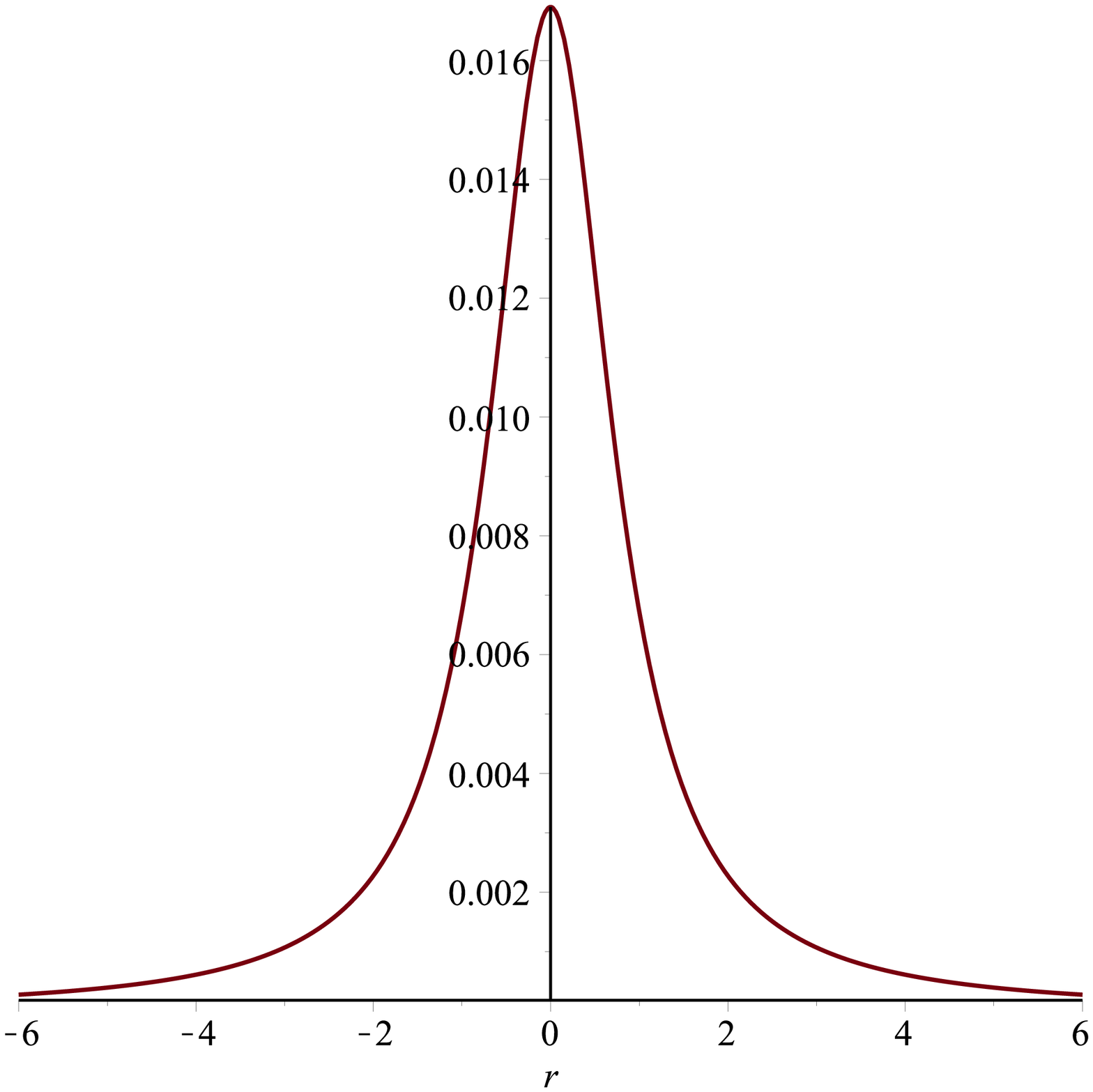}
\hss} \caption{\small Massless wormhole potential for null geodesics
with one maximum ($ m=0,\,l=0.1,\,n=1,\, b=1.3$).} \label{F9}
\end{minipage}
\hfill
\begin{minipage}[t]{0.48\linewidth}
\hbox to\linewidth{\hss%
  \includegraphics[width=0.95\linewidth,height=0.7\linewidth]{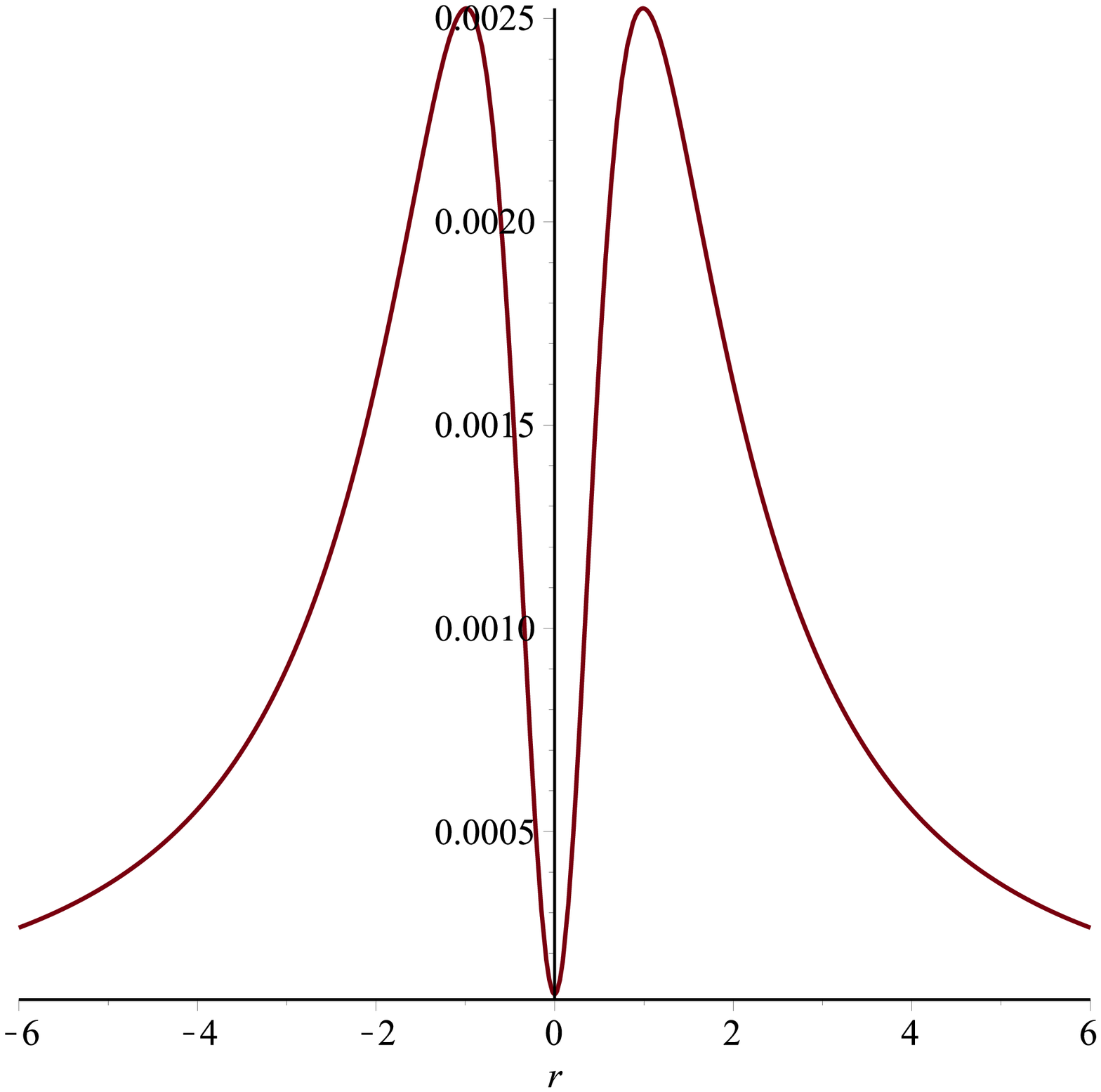}
\hss} \caption{\small Massless wormhole potential for null geodesics
with three extrema ($m=0,\,l=.1,\,n=1,\, b=0.1$). } \label{F10}
\end{minipage}
\end{center}
\end{figure}

In the case of timelike geodesics ($\varepsilon=-1$), for $m\neq 0$
the potential is asymmetric under reflection $r\to -r$. We will
assume $m > 0$ which corresponds to a positive mass from the point
of view of an observer at $r=+\infty$ and negative mass for an
observer at $r=-\infty$. In the generic case one maximum is at
negative $r$ while at $r>0$ there are two minima with a local
maximum between, or only one minimum. A typical potential with four
extrema is shown in Figs. 1-2. The case with only two extrema is
illustrated in Fig. 3. Remarkably, the potential for zero orbital
momentum $l=0$, generically has two extrema: a minimum at $r=r_+$
and a maximum at $r=r_-$ (see Fig. 4),
 \be\lb{extrl0}
r_\pm = \frac{e^2-2n^2 \pm \sqrt{e^4-4n^2b^2}}{2m}\,,
 \ee
corresponding to stable or unstable equilibrium positions. The
potential for the massless wormhole $m=0$ is symmetric under
reflection $r\to -r$, and so has an extremum at $r=0$, and two other
extrema at $r=\pm r_0$:
 \be
r_0^2 = \frac{(2b^2-n^2)l^2 + n^2 (b^2-n^2)}{n^2-b^2-l^2}\,, \ee
provided $0<n^2-b^2<l^2$ and $l^2 \le n^2(n^2-b^2)/(2b^2-n^2)$ (if
$4b^2>n^2/2$).

The potential for null geodesics $\varepsilon =0$ is positive
definite (Figs. 8-10). It has either one, or three extrema. In the
massless case $m=0$ there is an extremum at $r=0$, and two other
extrema at $r = \pm r_1$ with
 \be
r_1^2 = n^2-2b^2\,,
 \ee
provided $n^2>2b^2$. In this case there is a stable circular photon
orbit at the wormhole throat $r=0$, with areal radius $n$, and
unstable photon orbits at $r = \pm r_1$. The purely radial motion
($l=0$) of the photons is free.

By following the procedure of \cite{kagra}, exact solutions for
$r(\lambda)$ can be obtained in terms of Weirstrass elliptic
functions. Here we will only discuss the qualitative radial motion
for timelike or null geodesics sampling the wormhole ($b^2 > 0$)
geometry, and consider special limiting cases of interest. We shall
work in the rotated coordinate frame in which the $z$ axis is along
$\vec J$, so that the polar angle $\theta$ is constant.

\subsection{Timelike geodesics ($\varepsilon=-1$)}

\subsubsection{Scattering states}
The effective potential is for $m>0$  negative for large positive
$r$ and positive for large negative $r$, and so presents at least a
maximum and a minimum, with $r_{\rm min} > r_{\rm max}$, and
possibly also a secondary maximum and a secondary minimum. A test
particle coming from infinity with $E^2\ge1$ will first be
accelerated in the potential well around $r_{\rm min}$, before
either being reflected (scattered) by the potential barrier or, if
its energy is high enough, being transmitted through the wormhole
neck $r=r_{\rm max}$ to the second asymptotic sheet $r \to -\infty$.

Combining this radial motion with the angular motion previously
discussed leads to a trajectory which spirals around the polar axis
$\vec{J}$ at a constant angular velocity $J$ with respect to
$\lambda$. The definition (\ref{mint}) of $\lambda$ together with
(\ref{rad2}) leads to the relation
 \be
d\lambda = \frac{dr}{(r^2+n^2)[E^2 - 1 - U(r)]^{1/2}}\,.
 \ee
Putting $\beta^2 = E^2-1$, it follows that the geodesic range of
$\lambda$ is given by the finite integrals
 \be\lb{rangerefl}
\Delta\lambda_{\rm refl} =
2\int_{r_0}^{\infty}\frac{dr}{(r^2+n^2)[\beta^2 - U(r)]^{1/2}}
 \ee
in the case of reflexion (with $r_0$ the turning point, $U(r_0) =
\beta^2$), or
 \be\lb{rangetrans}
\Delta\lambda_{\rm trans} =
\int_{-\infty}^{\infty}\frac{dr}{(r^2+n^2)[\beta^2 - U(r)]^{1/2}}
 \ee
in the case of transmission (if $\beta^2 > \beta_0^2 = U_{\rm
max}$). Defining as usual the impact parameter by $a=l/\beta$, the
cross-section for transmission is
 \be
\sigma_{\rm trans} = \pi a_0^2 = \frac{\pi l^2}{U_{\rm max}}\,.
 \ee
In the limiting intermediate case $\beta^2 = \beta_0^2$, the
integral (\ref{rangerefl}) with the lower bound $r_{\rm max}$ is
logarithmically divergent, and the trajectory, attracted by the
unstable circular orbit $r=r_{\rm max}$, spirals indefinitely
towards the wormhole neck. In the present paper, we will only
evaluate these integrals for two special cases of interest: the
non-relativistic limit, and the small-NUT limit.

The non-relativistic limit is defined by $m=0$ and $b^2=n^2$ (so
that $f(r)=1$) and $\beta^2 \ll 1$. In this case,
 \be\lb{Unr}
U(r) = l^2/(r^2+n^2)
 \ee
leads to $r_0^2 = a^2 - n^2$, with
 \be\lb{rangereflnr}
\Delta\lambda_{\rm refl} =
\frac2\beta\int_{r_0}^{\infty}\frac{dr}{[(r^2+n^2)(r^2-r_0^2)]^{1/2}}
= \frac2lK\left(\frac{n}{a}\right)\,,
 \ee
where $K$ is the complete elliptic integral of the first kind. In
the frame $\psi=0$, the corresponding total angular variation is
 \be\lb{angrefl}
\Delta\varphi_{\rm refl} = J\Delta\lambda_{\rm refl} \simeq
\frac2{\sin\theta}K\left(\frac\beta2\cot\theta\right)\,,
 \ee
where we have used $E\simeq 1$ for small $\beta$. If $l \simeq
2n\tan\theta$ is held fixed in the non-relativistic limit
(corresponding to large impact parameters $a$), (\ref{angrefl})
reduces (using $K(0) = \pi/2$) to the finite result
$\Delta\varphi_{\rm refl} \simeq \pi/\sin\theta$. If on the other
hand the impact parameter $a$ is held fixed (of the order of $n$),
then the cone angle $\theta$ goes to zero, and the test particle
spirals around the polar axis many times before being scattered back
to infinity.

Similarly, in the case of transmission through the wormhole ($a <
a_0=n$),
 \be
\Delta\lambda_{\rm trans} =
\frac2\beta\int_{0}^{\infty}\frac{dr}{[(r^2+n^2)(r^2+n^2-a^2)]^{1/2}}
= \frac2{\beta n}K\left(\frac{a}{n}\right)\,.
 \ee
Again, if $\beta \ll 1$ the cone angle $\theta \simeq \beta a/2n$ is
very small, and the test particle makes a finite but large number of
turns
 \be\lb{turntrans}
\frac{\Delta\varphi_{\rm trans}}{2\pi} \simeq \frac2{\pi\beta}
K\left(\frac{a}{n}\right)
 \ee
along its path through the wormhole.

Remarkably, the number of turns remains large,
 \be\lb{turntrans0}
\frac{\Delta\varphi_{\rm trans}}{2\pi} \simeq \frac1\beta
 \ee
in the limit of a purely radial trajectory, $l=0$ ($a=0$). This
fact, which results from the conservation of total angular momentum
(\ref{J}) ($\vec{J} = \vec{S}$ in the case of purely radial motion,
$\vec{L}=0$), is of course not possible for a classical point test
particle, but makes sense if one considers rather a small extended
solid test body. For instance, (\ref{turntrans0}) means that a
tennis ball, of radius small before the wormhole throat radius $n$,
will spin around, with an angular velocity
 \be
\frac{d\varphi}{d\tau} = \frac{2nE}{r^2+n^2}
 \ee
which is maximum at the wormhole throat $r=0$, making $\beta^{-1}$
turns while going straight through the wormhole from $r=+\infty$ to
$r=-\infty$.

Let us emphasize that this uniform rotation of an extended test body
in free radial fall does not depend on the direction of this radial
motion, which generically does not coincide with that of the Misner
string. From (\ref{theta2}) and (\ref{solvarphi1}), the angular
motion along a parallel $\cal C$ is a uniform rotation, with proper
angular velocity $J/(r^2+n^2)$. In the limit of a purely radial
motion, $l\to 0$, the parallel $\cal C$ contracts to a point
$\theta=$ constant, $\varphi=$constant, while the uniform rotation
velocity retains the finite value $2nE/(r^2+n^2)$.

The small-NUT limit is motivated by the non-observation of NUT
charge in nature. It is defined by $n^2 \ll m^2$ and $n^2 \ll b^2$,
so that $\alpha^2 \simeq e^2$. In this limit the effective potential
(\ref{Upot}) can be approximated by
 \be\lb{pottsn}
U(r) \simeq \frac{-2mr^3 + (e^2+l^2)r^2 - 2ml^2r +
e^2l^2}{(r^2+n^2)^2},
 \ee
and its derivative by
 \be
U'(r) \simeq \frac{2r[mr^3 - (e^2+l^2)r^2 + 3ml^2r -
2e^2l^2]}{(r^2+n^2)^3}.
 \ee
Obviously $r = 0$ is a maximum of $U(r)$, and an absolute maximum
because $U(0)-U(r)\ge0$ in the small NUT limit:
 \be
\beta_0^2 =   U_{\rm max} \simeq U(0) \simeq
\left\vert\begin{array}{cc}\dfrac{e^2l^2}{n^4} & (l\neq0)\,, \\
\dfrac{e^2}{n^2} & (l=0)\,. \end{array}\right.
 \ee
We first consider the case of reflexion ($\beta^2 < \beta_0^2$),
which we will discuss only in the small-velocity limit
($\beta\to0$). In this limit $l=\beta a$ can be neglected (purely
radial motion), so that
 \be\lb{beta0}
\beta^2 - U \simeq -U \simeq \frac{2mr - e^2}{r^2+n^2}\,,
 \ee
with $r_0 \simeq e^2/2m$. The evaluation of the integral
(\ref{rangerefl}) yields a small, but non-vanishing
 \be
\frac{\Delta\varphi_{\rm refl}}{2\pi} \simeq \frac{2n}{e}\,.
 \ee

Transmission through the wormhole neck occurs if $\beta^2 >
\beta_0^2$, i.e. in the extreme-relativistic case. The cone angle is
given in this case by $\tan\theta \simeq a/2n < n/2e$. In the
extreme-relativistic limit we can approximate
 \be
\beta^2 - U \simeq \beta^2 - \frac{\beta_0^2n^4}{(r^2+n^2)^2} =
\beta^2\left[1 - \frac{e^2a^2}{(r^2+n^2)^2}\right]\,.
 \ee
The evaluation of the integral (\ref{rangetrans}) yields
 \be\lb{angtrans}
\Delta\varphi_{\rm trans} \simeq \dfrac{4n}{\sqrt{n^2+ea}}
K\left(\sqrt{\dfrac{2ea}{n^2+ea}}\right)\,.
 \ee
For very small impact parameters (corresponding to finite anglar
momenta $l = \beta a$) $a \ll n^2/e$, (\ref{angtrans}) reduces to
the finite result $\Delta\varphi_{\rm trans} \simeq 2\pi$. Again,
this remains valid in the limit of a purely radial trajectory, so
that a small extended test body falling straight through the
wormhole will at the same time spin around (at a maximum proper
angular velocity $d\varphi/d\tau \simeq 2\beta/n$), making exactly
one turn while going from $r=+\infty$ to $r=-\infty$.

\subsubsection{Bound states}
Bound states in the wormhole Brill geometry have some unusual
features with respect to the typical black hole situation. The
generic potential (Fig.~1) for massive particles has two potential
wells: one is at negative $U$ in the far region with small binding
energies (non-relativistic Newtonian orbits) shown in Fig. 2.
Another, relativistic potential well closer to the throat, extends
to positive energies, being separated from escape by a potential
barrier (Fig. 1) with larger binding energies. One has therefore
oscillating bound orbits between the two turning points of two
types: Newtonian, and relativistic. At the corresponding minima one
has two stable circular orbits. If two roots of the equation
$U'(r)=0$ are complex, the potential has a simpler shape with only
two extrema, Fig. 3, the minimum (potential well) being in the $r>0$
region, and the maximum at some  $r<0$. As it is seen from this
figure, the potential well is deep and relativistic, the Newtonian
well being absent in this case.

For zero orbital momentum $l=0$, the potential reduces to the simple
form
 \be
U(r) = \frac{(r-m)^2+b^2}{\rho^2}  - 1
 \ee
with a minimum and a maximum given by Eq. (\ref{extrl0}). One can
observe that the the shape of the potential for $l=0$ (Fig. 3) is
qualitatively similar to that for $l\neq0$ with two extrema (Fig.
4). Indeed for $l=0$ we still have a non-zero total angular momentum
$J=2nE$. The bound orbits are oscillating between two turning points
with relativistic velocities. The test particle can remain at rest
at the minimum at $r=r_+$ provided its squared energy
 \be
E_0^2 = \frac{e^2 - \sqrt{e^4-4n^2b^2}}{2n^2}
 \ee
is positive, i.e. if $b^2>0$. However for $n\neq0$, a small test
body at the equilibrium position $r=r_0$ will spin around with
angular velocity $\omega_0=2nE_0/(r_0^2+n^2)$. In the small-NUT
limit, these $l=0$ equilibrium values reduce to
 \be
r_0 \simeq \frac{e^2}m\,, \quad E_0 \simeq \frac{b}e\,, \quad
\omega_0 \simeq \frac{2nbm^2}{e^5}\,.
 \ee

The potential is particularly simple and reflection-symmetric in the
case of the massless wormhole $m=0$. Depending on the parameter
values, there could be three different cases: a potential well
centered at the throat extending to positive energy values and
separated from the scattering region by potential barriers (Fig. 5),
a pure potential barrier centered at the throat (Fig. 6) and a
purely negative potential well at the throat (Fig. 7). In the first
and the last cases one thus has stable bound orbits oscillating
around the throat, including circular ones located exactly at the
throat.

Another simple case is the  near-extreme limit $b^2 \ll m^2$. The
effective potential (\ref{Upot}) has its absolute minimum at $r=m$
(the extreme RN horizon radius) for $b^2=0$ ($e^2=m^2+n^2$). So for
$b^2$ small and positive, there are stable circular orbits at $r = m
+ {\rm O}(b^2)$, with energy $E_l$ and proper angular velocity
$\omega_l=d\varphi/d\tau$
 \be
E_l^2 \simeq b^2 \,\frac{e^2+l^2}{e^4}\,, \quad \omega_l \simeq
\frac{l}{e^2}\,.
 \ee

\subsection{Null geodesics ($\varepsilon=0$)}

The effective potential for null geodesics
 \be\lb{Unull}
U(r) = \frac{l^2[(r-m)^2 + b^2]}{(r^2+n^2)^2}\,,
 \ee
is shown in Figs. 8-10. The discussion of the scattering of light
parallels that for timelike geodesics. Of special interest is the
scattering angle $\Delta\varphi_{\rm refl} = J\Delta\lambda_{\rm
refl}$, where $\Delta\lambda_{\rm refl}$ is given by
(\ref{rangerefl}) with $\beta = E$, and $J^2 = (4n^2+a^2)E^2$. This
can be computed analytically in the special case $m=0$, with the
result
 \ba
\Delta\varphi_{\rm refl} &=& 2\sqrt{\frac{4n^2+a^2}{\alpha^2}}
K\left(\sqrt{1-\frac{r_0^2}{\alpha^2}}\right) \quad {\rm
for} \quad a^2 > \frac{n^4}{b^2}\,, \\
\Delta\varphi_{\rm refl} &=& 2\sqrt{\frac{4n^2+a^2}{r_0^2}}
K\left(\sqrt{1-\frac{\alpha^2}{r_0^2}}\right) \quad {\rm for} \quad
n^2 > 2b^2 \; {\rm and} \; 4(n^2-b^2) < a^2 < \frac{n^4}{b^2}\,,\nn
 \ea
with
 \be
\alpha^4 = a^2[a^2 + 4(b^2-n^2)]\,, \quad r_0^2 = \frac{\alpha^2+
a^2-2n^2}2\,.
 \ee
In the non-relativistic case $b^2=n^2$, this reduces for $a>n$ to
 \be\lb{scattnr}
\Delta\varphi_{\rm refl} =
2\sqrt{1+\dfrac{4n^2}{a^2}}\,K\left(\dfrac{n}{a}\right) =
\frac2{\sin\theta}K\left(\frac12\cot\theta\right) \,,
 \ee
where $\tan\theta > 1/2$, while light rays with small impact
parameters $a < n$ ($\tan\theta < 1/2$) end up in the second
asymptotic sheet ($r\to-\infty$) with
 \be
\Delta\varphi_{\rm trans} =
\frac4{\cos\theta}K\left(2\tan\theta\right)\,.
 \ee

Let us also note the existence, in the near-extreme limit $b^2 \ll
m^2$, of circular null orbits at $r \simeq m$ (as in the case of
massive particles) with energy
 \be
E_l^2 \simeq b^2 \,\frac{l^2}{e^4}\,.
 \ee

\setcounter{equation}{0}
\section{Charged particle motion}
Consider now the non-geodesic motion of a charged particle with mass
$m_c$ and charge $q_c$. It is convenient to work with the covariant
Lorentz equations of motion:
 \be
 \left(\dot{x}^\nu g_{\mu\nu}\right)\dot{} -
 \frac12 \partial_\mu g_{\nu\lambda} \dot{x}^\nu \dot{x}^\lambda
 = \kappa F_{\mu\nu} \dot{x}^\nu\,,
 \ee
where $\kappa=q_c/m_c$. Since $A_\mu$ has only $\mu=a=(t,\,\varphi)$
components depending only on $r,\,\theta$,  one has for $\mu=a$
 \be
F_{a\nu} \dot{x}^\nu=- \dot{A}_a  \,,
 \ee
while for the remaining $\mu=i=(r,\,\theta )$ components
 \be
F_{i\nu} \dot{x}^\nu = \partial_i A_a \dot{x}^a  \,.
 \ee
The equations of motion thus split into two groups. The first group
of equations do not contain the metric derivative term and combine
to a total derivative
 \be
 \left(\dot{x}^\nu g_{a\nu}+\kappa A_a\right)\dot{}=0 \,,
 \ee
 which can be integrated as before by introducing two integrals of motion
 $E,\, J_z$
  \ba
&& f(r)(\dot{t} - 2n(\cos\theta+C)\,\dot\varphi) - \kappa \Phi(r) =
E\,,\lb{Ech}\\ &&\dot{x}^\nu g_{\varphi\nu}+\kappa
A_\varphi=J_z+2nC\,,
 \ea
the second one being equivalent to (\ref{varphi}). In the second
group it is enough to consider only the $\theta$ component, taking
the normalization condition (\ref{rad}) as a constraint. The
$i=\theta$ component reads
 \be
[(r^2+n^2)\dot\theta]\,\dot{}=\frac12
g_{ab,\theta}\dot{x}^a\dot{x}^b +\kappa A_{a,\theta}\dot{x}^a =
g_{t\varphi,\theta}\dot{t}\dot{\varphi}+\frac12
g_{\varphi\varphi,\theta} \dot{\varphi}^2 +\kappa
A_{\varphi,\theta}\,\dot\varphi\,.
 \ee
Eliminating $\dot{t}$ from (\ref{Ech}), we find that the
$\kappa$-terms cancel, and we are left precisely with the equation
 \be
  [(r^2+n^2)\dot\theta]\,\dot{} = (r^2+n^2)\sin\theta\cos\theta
\,\dot\varphi^2 - 2nE\sin\theta\,\dot\varphi\,,
 \ee
identical to (\ref{theta0}). So the angular equations of motion are
the same as in the case of a neutral particle, leading to orbits
which lie on a ``cone'' of half-angle $\arctan(2nE/l)$ (with $l$ the
orbital angular momentum), while in the equation for the time
evolution (\ref{t}) and in the radial equation (\ref{rad1}) the
constant $E$ must be replaced by $E + \kappa\Phi(r)$ leading to
 \be\lb{rad3}
\dot{r}^2 + f(r)\left[\frac{l^2}{r^2+n^2}+1\right] =(E +
\kappa\Phi(r))^2 \,,
 \ee
where we took $\varepsilon=-1$ for massive, charged particles. This
can be rewritten as \be\lb{U} \dot{r}^2 + W(r)=0,\,\quad{\rm
where}\quad W(r)= f(r)\left[\frac{l^2}{r^2+n^2}+1\right] -(E +
\kappa\Phi(r))^2\,.
 \ee
The quantity $W(r)$ however is not convenient for the role of the
effective radial potential since it depends quadratically on $E$ and
contains unphysical negative energies. To correctly introduce the
radial potential we have to present this equation in the factorized
form
 \be\lb{rad4}
\dot{r}^2   =(E -V_+(r))(E-V_-(r)) \,,
 \ee
 where
  \be\lb{Vpm}
V_\pm(r)=-\kappa\Phi \pm \sqrt{f\left(1+\frac{l^2}{r^2+n^2}\right)}
\,,
 \ee
so that for a particle at rest $\dot{r}=0$ the correct branch
is\footnote{Recall that in special relativity the total energy of a
charge in the potential $\varphi$ (note that $\Phi$ is defined with
an opposite sign to $\varphi$) satisfying $(E-e\varphi)^2=m^2+{\bf
p}^2$ has to be solved with the sign plus:
$E=e\varphi+\sqrt{m^2+{\bf p}^2}$. In our case the mass is absorbed
by the affine parameter on the worldline, so $m=1$.} $E=V_+$.
Moreover, the quantity $E + \kappa\Phi(r)$ is equal to the kinetic
energy, it has therefore to be strictly positive (recall that $f>0$
everywhere for the wormhole):
 \be\lb{kinetic}
E + \kappa\Phi(r) =
\sqrt{\dot{r}^2+f(r)\left[\frac{l^2}{r^2+n^2}+1\right]}>0\,.
 \ee
It follows that $E>V_-$ always, so the particle's motion corresponds
to $E>V_+$. Also, the conditions for circular orbits $W=0=W'$ are
implied by the more physical conditions  $V_+=V_+'=0$ which do not
contain unphysical negative energies. So we can consider $V_+$ as a
correct radial potential.

To compare with the results of \cite{grunau} for the motion of
charged particles in the dyonic Reissner-Nordstr\"om spacetime, one
must first gauge transform our potential $\Phi(r)$ to a potential
vanishing at infinity $\Phi_0(r)$. This transformation and the
associated shift in the integration constant $E$ are
 \be
\Phi(r) = \Phi_0(r) + \frac{p}{2n}\,, \quad E = E_0 - \frac{\kappa
p}{2n}\,.
 \ee
Taking the NUT charge to zero, we find that the orbit of a charged
particle in the dyonic Reissner-Nordstr\"om spacetime lies on a cone
of half-angle $\arctan(\kappa p/l)$, in agreement with the results
of \cite{grunau}.

\subsection{Motion in the case $f=1$}
We restrict to the particularly simple case of the massless wormhole
with only magnetic and NUT charges related so tha $f(r)=1$,
 \be\lb{paraps2n}
m=0\,,\quad q=0\,,\quad p=\sqrt2n\,,
 \ee
leading to
 \be
V_+=-\frac{\kappa}{\sqrt{2}}\;\frac{r^2-n^2}{r^2+n^2}+\sqrt{
1+\frac{l^2}{r^2+n^2}} \,.
 \ee
The potential $V_+$ is symmetrical under reflection $r\to -r$. For
$\kappa>0$ it is maximum at the wormhole throat $r=0$ and
monotonically decreases with growing $|r|$ (Fig.~11). The maximal
value depends on the orbital momentum as follows:
 \be\lb{V0}
V_0 \equiv
V_+(0)=\frac{\kappa}{\sqrt{2}}+\sqrt{1+\frac{l^2}{n^2}}\,.
 \ee

\begin{figure}[tb]
\begin{center}
\begin{minipage}[t]{0.48\linewidth}
\hbox to\linewidth{\hss%
  \includegraphics[width=0.95\linewidth,height=0.7\linewidth]{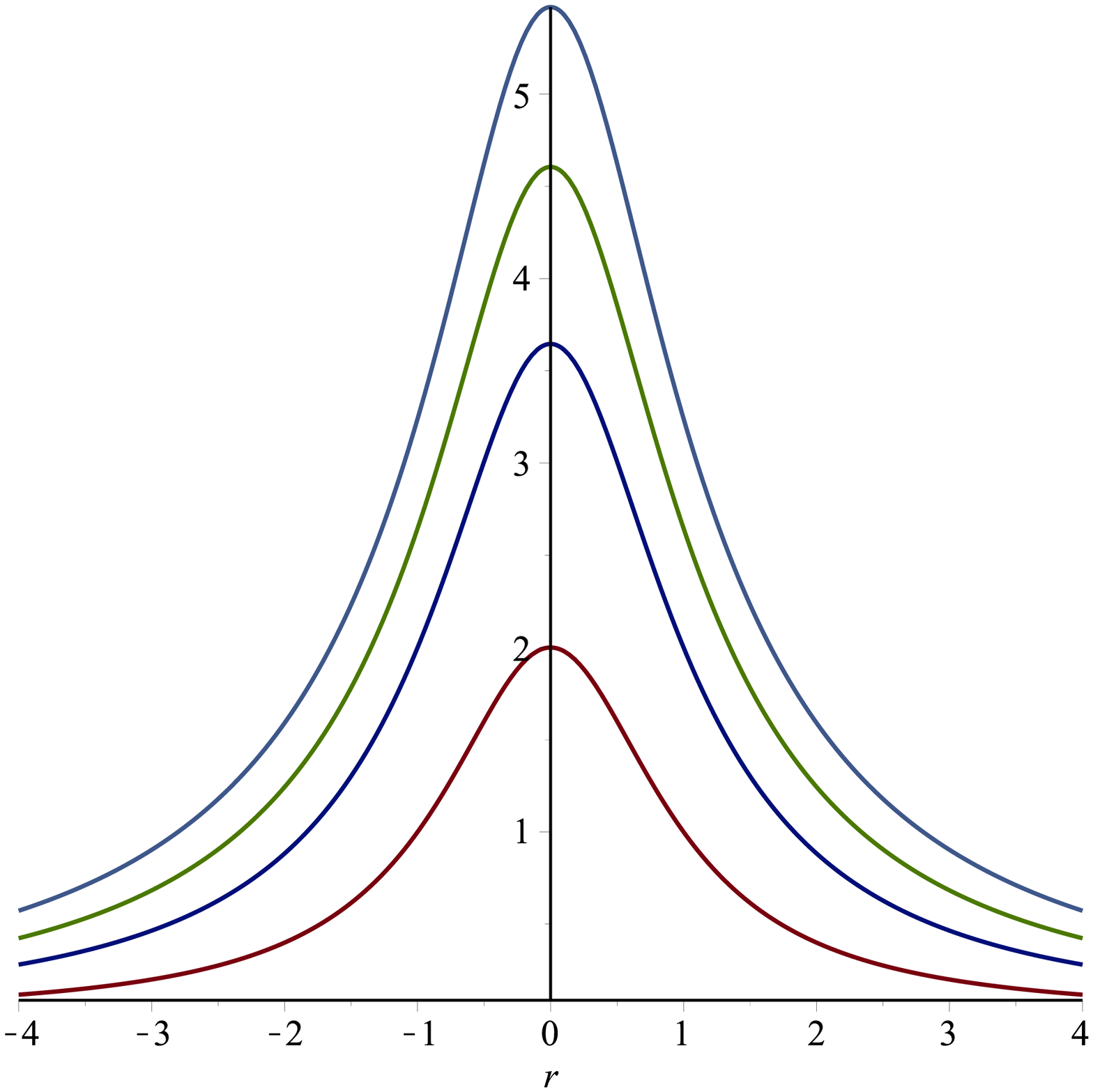}
\hss} \caption{\small Radial potential for charged particle motion
in the wormhole with $m=0,\;q=0,\;p^2=2n^2$ in the case of positive
$\kappa$ for different angular momenta $l^2$ ($\kappa=\sqrt{2},\;
l^2=19,\;12,\;6,\;0$ decreasing from top to bottom).} \label{F11}
\end{minipage}
\hfill
\begin{minipage}[t]{0.48\linewidth}
\hbox to\linewidth{\hss%
  \includegraphics[width=0.95\linewidth,height=0.7\linewidth]{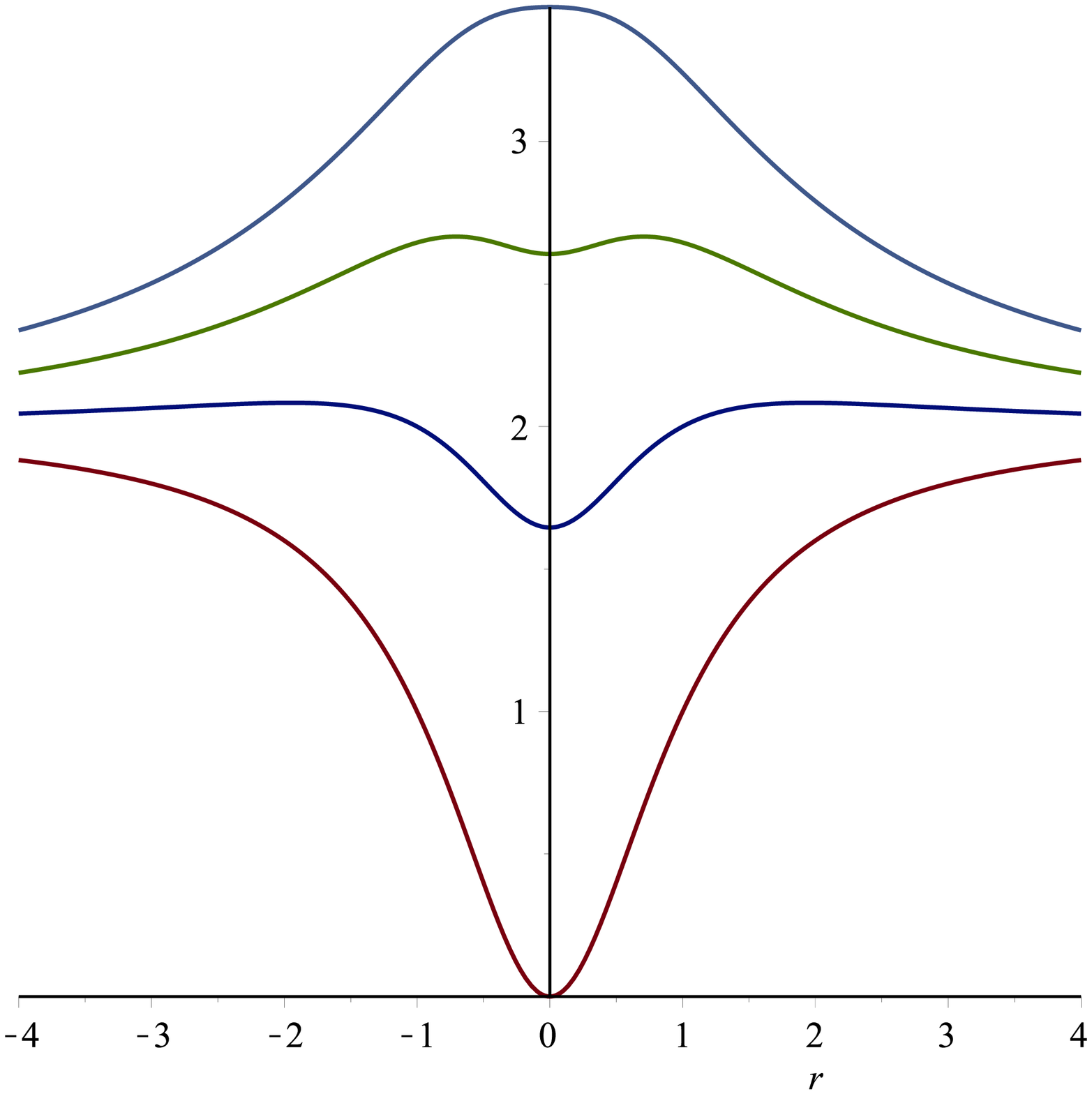}
\hss} \caption{\small Radial potential for negative
$\kappa\;(=-\sqrt{2})$ and the same values of $l^2$ from top to
bottom.} \label{F12}
\end{minipage}
\end{center}
\end{figure}

For $\kappa<0$ the shape of the potential is more diverse, as shown
in Fig.~12. In the region I: \be
 \frac{l^2}{n^2} <  2\sqrt2|\kappa|\,,
\ee  the potential has a local minimum at $r=0$. With growing $l^2$
the local minimum goes up and in the region II:
  \be\lb{condrpm}
 2\sqrt2|\kappa| < \frac{l^2}{n^2} < 4\kappa^2 + 2|\kappa|\sqrt{2
(1+2\kappa^2)}\,,
 \ee
local maxima develop on both sides of the throat, with the positions
 \be\lb{r0pm}
r_\pm= \pm n\,\sqrt{\frac{8\kappa^2 n^2(n^2+l^2)-l^4}{l^4-8\kappa^2
n^4}}\,.
 \ee
Finally, in the region III, \be
 \frac{l^2}{n^2} > \left(4\kappa^2 + 2|\kappa|\sqrt{2
(1+2\kappa^2)}\right) \,,
 \ee
the two local maxima at $r=r_\pm$ disappear, while the local minimum
at $r=0$ turns into an absolute maximum.

With the potentials shown in Fig.~11,~12 one deduces the following
characterization of the orbits. For $\kappa>0$ one has orbits
traversing the wormhole throat for $E>V_{0}$ corresponding to the
top of the potential barrier in Fig.~11. The orbits with $E<V_{0}$
correspond to scattering on the wormhole. For $\kappa<0$ in the
parameter region III the situation is the same. In region II one has
in addition bound states between the turning points inside the
potential well between the local maxima. The same energies
correspond to scattering orbits reflected on the exterior sides of
the barriers. The region I contains bound orbits inside the well and
traversing orbits for the energies above the wells. These can be
analyzed in analogy with Sec.~4, we leave it for future work.

The circular orbits at the throat $r=0$ (with areal radius $n$)
exist for both signs of $\kappa$. These will be interesting for
analysis of non-causality in the next section. They are unstable for
positive $\kappa$ and negative $\kappa$ in the region III, but
stable for negative $\kappa$ in the regions I and II. The
corresponding gauged energy $E=V_+(0)+\kappa/\sqrt2$ is positive for
positive $\kappa$, but it is negative for negative $\kappa$ if
 \be
\frac{l^2}{n^2}<2\kappa^2-1 \qquad (|\kappa| > 1/\sqrt2)\,.
 \ee
In the case of negative $\kappa$, two other unstable circular orbits
exist in region II at $r=r_\pm$  corresponding to the kinetic energy
 \be\lb{kinrpm} E+\kappa\Phi(r_\pm)=\frac{l^2}{2\sqrt{2} |\kappa| n^2}\,, \ee
and positive total energy
 \be\lb{Ecirc}
E=\frac{|\kappa|}{\sqrt{2}} \left(\frac{l^2}{4\kappa^2
n^2}+1+\frac{2n^2}{l^2}\right)\,. \ee

\setcounter{equation}{0}
\section{Causality}
The ADM form of the metric(\ref{nworm}) is
 \be\lb{adm}
ds^2 = - \frac{f(r^2+n^2)\sin^2\theta}{\Sigma}\,dt^2 + f^{-1}dr^2 +
(r^2+n^2)d\theta^2 +
\Sigma\left(d\varphi+\frac{2nf(\cos\theta+C)}{\Sigma}\,dt\right)^2,
 \ee
with
 \be
\Sigma(r,\theta) = (r^2+n^2)\sin^2\theta - 4n^2f(\cos\theta+C)^2\,.
 \ee
For $f(r)>0$ (which is always the case for the wormhole Brill
solution with $b^2>0$), $\Sigma$ becomes negative, and closed
timelike curves (CTCs) appear, in a neighborhood of the Misner
string given by $\Sigma(r,\theta)<0$. The surface
$\Sigma(r,\theta)=0$ bounding this CTC neighborhood is a causal
singularity of the spacetime, where the signature of the spacetime
changes from $(-+++)$ outside to $(+++-)$ inside. This singularity
is, just as the Misner string itself, completely transparent to
geodesic motion. Nevertheless, the occurrence of CTCs in a spacetime
is usually considered to violate causality \cite{hawking,thorne}. An
observer traveling around such a CTC would eventually return to his
original spacetime position after a finite proper time lapse, thus
opening the possibility for time travel. However, unless this
observer is freely falling, such a CTC travel would necessarily
involve accelerations generated e.g. by rocket engines. One can
argue that the back-reaction of these matter accelerations on the
spacetime geometry will deform it in such a way that chronology will
ultimately be preserved. If this reasoning is correct, in vacuum
gravity causality violation can only occur in spacetimes with closed
timelike geodesics (CTGs), or possibly closed null geodesics (CNGs).
However we deal here with a charged solution, so we must also
consider the worldlines of charged particles.

\subsection{No closed timelike or null geodesics}
We now show that there are no closed timelike or null geodesics in
the Brill spacetimes with $|C|\le1$. Closed geodesics (or
self-intersecting geodesics) occur if after a finite lapse of affine
parameter $\tau$ or of Mino time $\lambda$, all the coordinates take
again the same values (modulo $2\pi$ for the azimuthal angle). So
the perimeter of the geodesic must be an integer multiple of the
$\lambda$-period $2\pi/J$ of the angular motion. From
(\ref{Dttheta}), during this period,
 \be\label{Dtang}
\Delta t_{\theta} \ge 4\pi n\left(\frac{2nE}J-1\right)
 \ee
for $|C|\le1$. Also, from (\ref{tr}) and (\ref{rad1}),
 \be
\frac{dt_r}{d\lambda} = \frac{E(r^2+n^2)}{f(r)} \ge E^{-1}[l^2 -
\varepsilon(r^2+n^2)] \ge E^{-1}[l^2 - \varepsilon n^2]\,,
 \ee
leading to
 \be
\Delta t_r \ge \frac{2\pi}{EJ}\left[l^2 - \varepsilon n^2\right]\,,
 \ee
over the same period. Adding the two together, we obtain
 \be
\Delta t = \Delta t_r + \Delta t_{\theta} \ge \frac{2\pi}E\left[J -
2nE - \varepsilon n^2/J\right]\,.
 \ee
For $\varepsilon = -1$ this is clearly positive definite. For
$\varepsilon = 0$, this can vanish only for $2nE = J$ ($l=0$). But
in this case $\Delta t_{\theta} \ge 0$, while $dt_r/d\lambda$, and
thus also $\Delta t_r$, is positive definite. Thus, for $|C|\le1$
all timelike or null geodesics are causal (future directed).

The situation for $|C|>1$ is less obvious. We shall only discuss the
case of null geodesics. As we shall show, 1) there are always CNGs
in any given Brill spacetime if $|C|$ is large enough; 2) for any
given value of $|C|>1$, there are Brill spacetimes with CNGs.

First, we observe that for $C>1$,
 \be\label{Dtang1}
\Delta t_{\theta} = 4\pi n\left[\frac{2nE}J - |C|\right]
 \ee
for orbits which circle the South Misner string ($J_z < -2nE$). The
same relation (\ref{Dtang1}) holds for $C<-1$ in the case of orbits
which circle the North Misner string ($J_z > 2nE$). We wish to show
that, if the orbit is closed, this negative angular contribution
($2nE<J$) can be exactly balanced by the positive radial
contribution, leading to $\Delta t=0$ and thus to a closed geodesic.
Depending on the values of the model parameters, two situations can
lead to closed orbits with $\varepsilon=0$. Either $U(r)$ has a
local minimum $U_{\rm min}>0$ between two local maxima, and there
are bound states in the resulting potential well, leading to closed
orbits if the periods of the radial and angular motion are
commensurate (i.e. for a dense set of values of the energy). Or
$U(r)$ has only a maximum, corresponding to an unstable circular
orbit. Let us discuss these two cases separately.

In the first case, from (\ref{tr}) and (\ref{rad1}),
 \be
\frac{dt_r}{d\lambda} = l^2\frac{E}{U(r)} \le l^2\frac{E}{U_{\rm
min}}\,.
 \ee
It then follows that, over an angular period,
 \be\lb{boundDt}
\Delta t \le 4\pi n\left[\left(1+\frac{l^2}{4n^2U_{\rm
min}}\right)\frac{2nE}J - |C|\right] < 4\pi n\left[1 +
\frac{l^2}{4n^2U_{\rm min}} - |C|\right]\,,
 \ee
vanishing for $|C| = 1 + l^2/4n^2U_{\rm min}$ (which depends only on
the wormhole parameters $(m,n,b)$). For higher values of $|C|$, the
bound state null geodesics will be past directed.

In the case of a circular orbit of radius $r_0$ and energy
$E_0=U(r_0)$ (circling the North or South Misner string as above),
 \be
\frac{dt_r}{d\lambda} = \frac{l^2}{E_0} =
\frac{J^2-4n^2E_0^2}{E_0}\,,
 \ee
leading to
 \be\lb{Dtnullcirc}
\Delta t = \frac{2\pi}{E_0}[J - 2nE_0|C|]\,,
 \ee
which vanishes for
 \be\lb{cngcirc}
|C| = \frac{J}{2nE_0} = \left(1+\frac{l^2}{4n^2U(r_0)}\right)^{1/2}
 \ee
(value depending only on the wormhole parameters). So again in this
case, for any given set of Brill parameters $(m,n,b)$, there will be
CNGs for large enough $|C|$.

We now formulate the converse, and stronger, proposition: that for
any given $|C| > 1$, one can find a set $(m,n,b)$ such that there
are CNGs. Presumably this also extends to CTGs. Actually this
proposition is easy to prove in the small NUT limit. There is an
unstable circular orbit at $r \simeq 0$, of energy $E_0 \simeq
l^2e^2/n^4$ for $l\neq0$. From the above, there is a family of
corresponding closed null geodesics for
 \be
|C| \simeq \left(1+\frac{n^2}{4e^2}\right)^{1/2} \quad
\Longrightarrow \quad |C| \simeq 1 + n^2/8e^2 \,,
 \ee
which proves the proposition.

\subsection{Closed charged worldlines}
Since we deal with a charged solution, it is natural to
also consider the worldlines of charged particles. In this case we
have
 \be
\Delta t_r \ge \frac{2\pi}{Jf}(E+\kappa \Phi)(r^2+n^2) \,,
 \ee
which is positive as in the geodesic case by virtue of
(\ref{kinetic}). The angular contribution $\Delta t_\theta$ still
satisfies Eq. (\ref{Dtang}) (restricting to $|C|\leq 1$) and can be
negative. Consider circular orbits at the wormhole throat $r=0$ in
the case $f(r)=1$, $q=0$, $p=\sqrt2n$. From (\ref{kinetic}), $\Delta
t_r = 2\pi n\gamma/J$, with $\gamma=\sqrt{l^2+n^2}$, so that the
total $\Delta t$ is equal to
 \be
\Delta t = \frac{2\pi n}{J}\left(\gamma+4nE-2J\right)\,,
 \ee
with $J=\sqrt{4n^2E^2+\gamma^2-n^2}$. Denoting
$\Delta=(\gamma+4nE)^2-4J^2$ and using $E=V_+(0)$ given by
(\ref{V0}) we obtain
 \be
\Delta=5(\gamma-\gamma_-)(\gamma-\gamma_-+)\,,\qquad
\gamma_\pm=-\frac{2\sqrt{2}\kappa n}{5}
\left(1\pm\sqrt{1-\frac{5}{2\kappa^2}}\right)\,.
 \ee
Since $\gamma$ must satisfy $\gamma > 1$, we find that $\Delta$ is
non-positive for
 \be\lb{cwl}
\kappa<-\frac{9}{4\sqrt{2}}\,,\qquad \gamma_- \leq \gamma \leq
\gamma_+ \,.
 \ee
Note that the upper bound (\ref{cwl}) is smaller than
$4\sqrt2|\kappa|n/5$, leading  to $l^2/n^2 < 32\kappa^2/25$, which
is much lower than the upper bound (\ref{condrpm}) of region II, so
that the circular worldlines $r=0$ satisfying (\ref{cwl}) are
stable. We expect that nearby bound worldlines in the potential well
above $r=0$ will similarly violate causality. However, such
worldlines are confined to a narrow region around the wormhole
throat, so that these causality violations are actually
unobservable, for an observer initially living at a large distance
$r_1$ from the wormhole.

Now let us argue that all the worldlines which can be followed by
such an observer returning to his starting point are causal. The
negative angular contribution to the total round trip time lapse
$\Delta t_{\rm tot}$ is given by the contribution (\ref{Dtang})
during one period multiplied by a finite number of periods
proportional to
 \be\lb{rangewl}
\Delta\lambda_{\rm refl} =
2\int_{r_0}^{r_1}\frac{dr}{(r^2+n^2)[(E-V+(r))(E-V_-(r))]^{1/2}}.
 \ee
This will be easily balanced by the positive radial contribution
which is proportional to the distance between the turning point
$r_0$ and $r_1$, so that the worldline will remain causal in the
large. The observer travelling in a charged rocket and returning to
his starting point will not meet his younger self.

However this reasoning fails if, for a given value of the orbital
angular momentum $l$, the traveller's energy is just a bit below the
maximum of $V_+(r)$. The number of $\lambda$-periods (\ref{rangewl})
diverges logarithmically for $r_0 = r_{\rm max}$ so that, for $r_0$
close to $r_{\rm max}$, the charged particle makes many turns around
the wormhole throat before returning towards infinity. Then, the
angular time lapse could be very large (and negative), and the
worldline could possibly self-intersect at multiple radii $r = r_i$,
if the attractor unstable worldline at $r = r_{\rm max}$ was closed.
But we have shown above that the wordlines at $r=0$ can be
closed only if they are stable. In the region II parameter range
(\ref{condrpm}), there are unstable circular orbits at $r=r_\pm$.
During a period, the total $\Delta t$ is
 \be
\Delta t = \frac{2\pi}{J}\left[(r_\pm^2+n^2)(E+\kappa
\Phi(r_\pm))+4n^2E-2n\sqrt{4n^2E^2+l^2}\right]\,,
 \ee
where $r_\pm$ is given by (\ref{r0pm}), $E+\kappa\Phi(r_\pm)$ by
(\ref{kinrpm}) and $E$ by (\ref{Ecirc}). To see whether the square
root term may dominate we calculate the difference of the squared
quantities
 \be [(r_\pm^2+n^2)(E+\kappa\Phi(r_\pm))+4n^2E]^2 - 4n^2(4n^2E^2+l^2) =
\frac{8\kappa^2n^4l^2}{(l^4-8\kappa^2n^4)^2}\left[l^6+2(l^2+4n^2)
(l^4-8\kappa^2n^4)\right]\,,
 \ee
which is positive by virtue of Eq. (\ref{condrpm}), so that these
unstable circular wordlines are causal. So, while this does not
constitute a rigorous proof, we believe that there are no closed
charged wordlines extending to spacelike infinity. The case of
wormhole parameters more generic than those (\ref{paraps2n})
considered here should similarly be investigated.

\setcounter{equation}{0}
\section{Physical aspects}
\subsection{Tidal accelerations and traversability}
In order to be traversable, macroscopic Lorentzian wormholes should
not only be geodesically complete, with a class of timelike or null
geodesics connecting the two asymptotic regions, but they should
also be such that the tidal gravitational forces exerted on an
observer travelling through remain reasonably small \cite{MT1}. We
will discuss here only the case of radial geodesics ($l=0$). In the
non-relativistic case ($f(r)=1$), the radial velocity is constant,
so the longitudinal tidal force will vanish. However we have noted
previously that an extended object falling radially through the
wormhole will rotate at a uniform proper angular velocity
 \be
\omega(r) = \frac{2nE}{r^2+n^2}\,.
 \ee
which should generate a centrifugal acceleration $a(\vec{r})$. In
flat space, this would be given in cylindrical coordinates $(\rho,
z)$ by
 \be
a_{\rm flat} = \ddot{\rho} = \omega^2(z)\,\rho \simeq
\frac{4n^2}{(r^2+n^2)^2}\,\rho\,,
 \ee
where we have assumed for simplicity that the direction of motion is
along the $z$ axis, approximated $r \simeq z$ (the radius of the
object should be small before the wormhole throat radius $n$), and
$E \simeq 1$ (non-relativistic limit). Taking as in \cite{MT1} the
reference values $\rho_{\rm max} = 2\,$m for the radius of the
object and $a_{\rm max} = g = 9.8\, {\rm ms}^{-2}$ (Earth gravity),
we arrive at the rough estimate at the throat $r=0$
 \be\lb{flatest}
\frac{\ddot{\rho}}{\rho}(0) \simeq \frac4{n^2} \sim \dfrac{a_{\rm
max}}{\rho_{\rm max}} \sim 5\,{\rm s}^{-2}\,,
 \ee
leading to a NUT charge $n \sim 1$s, corresponding to a wormhole
throat radius $n \sim 3.10^8$m (or in mass units,
$n\sim 10^5 M_{\odot}$).

To improve on this flat-space estimate, we must compute the tidal
acceleration between two neighboring points of the freely falling
test body with purely spatial separation $\delta x$ in the comoving
frame \cite{MT1}, i.e. such that
 \be\lb{spatsep}
u_\mu\delta{x}^\mu = 0\,.
 \ee
This tidal acceleration is given by the geodesic deviation equation
\cite{MTW}
 \be\lb{geodev}
\delta\ddot{x}^\mu \equiv \dfrac{D^2}{d\tau^2}\delta{x}^\mu = -
{R^\mu}_{\rho\nu\sigma}u^\rho\delta{x}^\nu u^\sigma\,.
 \ee
Before proceeding, we note that in the non-relativistic case the
length scale is set by the sole NUT parameter $n$, so that the
Riemann tensor components at the throat, and hence the relative
tidal accelerations, must be of order $n^{-2}$, as in our flat-spacet
estimate.

In the case of radial geodesics the four-velocity $u^\mu =
\dot{x}^\mu$ has only two non-vanishing components $u^0$ and $u^1$,
so that the condition (\ref{spatsep}) reduces to
 \be
(g_{00}\delta{t} + g_{03}\delta{\varphi})\,u^0 +
g_{11}\delta{r}\,u^1 = 0\,,
 \ee
which may be used to eliminate $\delta{t}$ from the right-hand side
of (\ref{geodev}). Taking into account the non-vanishing values of
the Riemann tensor components listed in Appendix A, we thus obtain
for the longitudinal and transverse accelerations
 \ba
\delta\ddot{r} &=& - \left[{R^r}_{ttr}\delta{t}u^0u^1 +
{R^r}_{trt}\delta{r}(u^1)^2 + {R^r}_{t\varphi
r}\delta{\varphi}u^0u^1\right] \nn\\
&=& \frac{(f-1)(3r^2-n^2) + 4mr}{(r^2+n^2)^2}\,\delta{r}\,,\lb{longacc}\\
\delta\ddot{\theta} &=& - \left[{R^\theta}_{t\theta t}(u^0)^2 +
{R^\theta}_{r\theta r}(u^1)^2\right]\,\delta\theta \nn\\
&=&
\frac{r(r-m)-(r^2-n^2)f}{(r^2+n^2)^2}\,\delta\theta\,,\lb{transacc}
 \ea
and a similar equation for $\delta\ddot{\varphi}$. In the
non-relativistic case $f=1$ (implying $m=0$), the longitudinal
acceleration (\ref{longacc}) vanishes identically, while the
transverse acceleration takes the value at the wormhole throat $r=0$
 \be
\delta\ddot{\theta}(0) = \frac{1}{n^2}\,\delta\theta\,.
 \ee
This is precisely one fourth of the flat space estimate
(\ref{flatest}), meaning that our traversability estimate for the
NUT charge should
be halved, $n \sim 0.5\,{\rm s} \sim 1.5 \times 10^8\,{\rm m}$. At
present we have no simple explanation for this factor $1/4$.

For generic wormhole parameters, the longitudinal and transverse
accelerations at the wormhole throat take the values
 \ba
\delta\ddot{r}(0) &=& - \frac{e^2-2n^2}{n^4}\delta{r}\,,\nn\\
\delta\ddot{\theta}(0) &=& \frac{e^2-n^2}{n^4}\delta{\theta}\,.
 \ea
The transverse (centrifugal) acceleration is always positive
($e^2-n^2 > m^2$), while the longitudinal acceleration can have
either sign. In the small-NUT limit, the two accelerations become
large (of order $e^2/n^4$) and opposite in sign, so that they work
together to compress the test body along the direction of infall.

\subsection{Electromagnetic fields}
In the present case, the Lorentzian wormhole geometry is generated
by electric and/or magnetic fields which will become large near the
wormhole throat. For simplicity we discuss only the case of purely
electric or purely magnetic monopole fields.
In Gaussian units these fields are given by
 \be\lb{fields}
\begin{array}{ccl}
E = & F_{01} & =  {q} \dfrac{r^2-n^2}{(r^2+n^2)^2} \quad (p = 0)\,,
\\B = & \dfrac1{\sqrt{|g|}}F_{23} & = p
\dfrac{r^2-n^2}{(r^2+n^2)^2} \quad (q = 0)\,,
\end{array}
 \ee
with $e^2=q^2+p^2>n^2+m^2$ for NUT wormholes. We first consider a
purely electric field, in which case $q>n$, so that the maximal
field is $E_{\rm max} > n^{-1}$. A first constraint to be satisfied
for the validity of the classical approximation which we have used
is that the Schwinger pair creation should be unobservable. In flat
space, the Schwinger critical field in $\hbar=c=1$ units reads
$E_{\rm Sch}=\mu_e^2/e_0$, where $\mu_e,\,e_0$ are the mass and the
charge of the electron. The ratio can be put into the following form
 \be
\frac{E_{\rm max}}{E_{\rm Sch}} = e_0\frac{M_{\odot}}{n}\frac{m_{
Pl}}{M_{\odot}} \left(\frac{m_{ Pl}}{\mu_e}\right)^2\,,
 \ee
where $m_{ Pl}$ is the Planck mass. Inserting here ${m_{
Pl}}/{M_{\odot}}=1.1\times 10^{-38},\;{m_{ Pl}}/{\mu_e}=2.4\times
10^{22},\; e_0=137^{-1/2}$ one obtains
 \be \frac{E_{\rm max}}{E_{\rm Sch}}=0.54\times 10^6 \,\frac{M_{\odot}}{n}\,. \ee
Thus in the purely electric case the classical picture is valid if
$n \gtrsim 5\times10^5\, M_{\odot}$ (which is of the order of the geometric radius
of the Sun  $ R_{\odot} \sim
7 \times10^8\,{\rm m}$). This lower bound for $n$, which should be
improved by considering pair creation in the wormhole spacetime
metric is only six times larger than our traversability estimate.

We should also consider the effect of large electric fields
on infalling test bodies. While ordinary matter is electrically
neutral, so that its motion is not affected by electromagnetic
fields, it is made up of charged atomic nuclei and electrons, which
are sensitive to these fields. For a spaceship to go through the
wormhole without damage, the electric field at the throat should be
smaller than the ionization threshold. One can take as
the relevant electric field the Coulomb field of the electron at the
Bohr radius
 \be
E_i=\frac{e_0}{r_B^2},\quad r_B=\frac1{\mu_e
e_0^2}\,.
 \ee
One finds in the same units:
 \be
\frac{E_i}{E_{\rm Sch}}=e_0^6=137^{-3}=0.4\times 10^{-6}\,.
 \ee
Therefore the threshold for traversability of an electric wormhole
by an unmanned spaceship is $n\gtrsim 10^{12}\, M_{\odot} \sim
10^{15}\,{\rm m} \sim 0.1$ lt-yr. Traversability by a manned
spaceship would impose even more drastic constraints. For instance,
molecular air is ionized by a constant electric field $E \sim 3
\times 10^{6}\,{\rm V}\,{\rm m}^{-1} \sim 10^{-6} \,E_i$; to avoid
this the throat radius $n$ should be at least equal to $10^5$ lt-yr,
of the order of the radius of our Galaxy, which seems totally
unrealistic.

In the case of a purely magnetic wormhole no pair creation occurs,
but above the Schwinger threshold one can expect phenomena of
superstrong magnetic field such as in certain neutron stars. For the
wormhole radius value $n \sim 1.5 \times 10^8\,$m derived in the
previous subsection, the wormhole-throat field values would be $B(0)
\sim 3.3 \times 10^{10}\,{\rm T}$. For comparison, human exposure to
a constant magnetic field of up to $B \sim 8\,{\rm T}$ can be
tolerated \cite{static}. Taking the latter (magnetic) value as
reference would lead to the lower limit for the wormhole radius $n >
6\times10^{17}\,{\rm m} \sim 60\,$lt-yr, corresponding to a huge NUT
charge, larger than our previous tidal force estimate by nine orders
of magnitude. However, traversability of a purely magnetic NUT
wormhole by an unmanned spaceship would seem to be feasible for not
too large wormhole radii.

\section{Outlook}
In this paper we have suggested a novel type of wormholes without
exotic matter. They are supported by matter sources satisfying the
NEC condition and evade the NEC-violation theorem due to the
presence of a non-diagonal ($t,\, \varphi$) component of the stress
tensor, associated with a NUT charge. The price to pay for this is
that the spacetime is not strictly asymptotically flat, but only
locally asymptotically flat, and presents a Misner string
singularity. The particular realization described here is supported
by a Maxwell field and corresponds to a supercritically charged
RN-NUT solution without a central singularity or horizons. Extending
the results of \cite{Clement:2015cxa}, we have analyzed in detail
the geodesic motion and shown that the Misner string is transparent
for geodesic motion without periodic identification of time, so that
the spacetime is geodesically complete. This result is independent
of the actual form of the gravitational potential $f(r)$, and holds
also for the black-hole and extreme black-hole RN-NUT solutions. A
curious effect of the NUT charge is that small freely falling test
bodies are endowed with a proper angular momentum or spin, which is
aligned with the direction of infall, and is independent of that of
the Misner string.

Even without periodic identification of time, the
RN-NUT spacetimes contain regions with closed time-like curves. We
have shown, however, that for some subfamily of these spacetimes there
are no closed timelike or null geodesics, so that freely falling
observers should not encounter causality violations. On the other
hand, the analysis of the motion of charged test particles has
revealed that there are wordlines which can be closed or become past
directed if their charge-to-mass ratio is large enough. This was
found, however, for orbits threading the throat and not reaching
asymptotic regions of space-time, so the existence of observable
manifestations of non-causality still remains open.

We have also investigated under what conditions these RN-NUT
wormholes are macroscopically traversable, and shown that the
gravitational tidal forces at the throat could be kept reasonably
small for a reasonably small NUT charge. However, the
electromagnetic forces acting on ordinary matter under the same
conditions would be extremely large, and could be kept under control
only in the case of huge NUT charges.

This work could be extended in several directions. First, our
analysis was purely classical. One should explore propagation of
waves and quantum implications of NUT wormholes, as well as their
stability \cite{Holzegel:2006gn}. Also, similar NUT wormhole
solutions could presumably be constructed in other gravitating field
theories. One such solution was found recently in the Einstein-Skyrme
theory \cite{Ayon-Beato:2015eca} after our paper was completed. Five-dimensional vacuum
gravity is known to admit NUTless Lorentzian wormhole solutions
\cite{Chodos:1980df}, which are however not traversable
\cite{AzregAinou:1990zp}. We suggest that the possible existence of
geodesically complete and traversable NUTty wormhole solutions to
this theory be investigated. Other five-dimensional theories, such
as five-dimensional minimal supergravity (which can be consistently
truncated to the four-dimensional Einstein-Maxwell theory
\cite{oxi}), should also presumably admit such solutions.

Moreover, by superposing NUT-anti-NUT configurations one could try
to construct asymptotically flat wormhole spacetimes. In the extreme
black-hole case ($b^2=0$), one can construct stationary
Israel-Wilson-Perj\`es NUT-anti-NUT superpositions. These are
singular if time is periodically identified \cite{Hartle:1972ya},
but can be regular (with only a Misner string connecting the two
event horizons) if it is not. From the present analysis, they should
be geodesically complete. It would be interesting to extend this to
the construction of slowly orbiting NUT-anti-NUT wormhole dynamical
solutions in the near-extreme case, i.e. for small $b^2>0$. These
should be asymptotically flat, with only a non-contractible Misner
string loop threading the two wormhole throats. If the parameters
could be arranged so that this two-wormhole configuration with two
regions at spacelike infinity presented a high degree of symmetry,
we speculate that it might be possible to identify these two
asymptotically flat regions \cite{misner2,lindquist,Clement95} to
yield a traversable Wheeler-Misner wormhole \cite{WM} connecting two
distant parts of a spacetime with only one region at infinity.

We conclude with some astrophysical speculations. Recently it was
suggested that wormholes could exist in the  galactic centers
 \cite{Li:2014coa,Rahaman:2014pba} or in the outer regions of halos
\cite{Rahaman:2013xoa,Kuhfittig:2013hva}. The option to have a
wormhole without exotic matter seems attractive:  its mass and NUT
charge could be reasonably of the scale of the mass concentration
in the galactic centers ($10^8 - 10^9 \, M_{\odot}$), in which case
it would not be destroyed by quantum effects. However, crucial
for this wormhole is the supercritical electric or magnetic charge,
whose origin in astrophysical conditions is unclear for the moment.
Nevertheless we think that further study of new realizations of
gravitational fields with NUTs may be useful for future
astrophysical applications.

\section*{Acknowledgments} DG and MG would like
to thank LAPTh Annecy-le-Vieux for hospitality at different stages
of this work. DG also acknowledges the support of the Russian
Foundation of Fundamental Research under the project 14-02-01092-a.
MG acknowledges the support  of the  Ministry of Higher Education
and Scientific Research of Algeria (MESRS) under grant 00920090096.

\renewcommand{\theequation}{A.\arabic{equation}}
\setcounter{equation}{0}
\section*{Appendix A: Symmetries}
The metric (\ref{nworm}) has the following four Killing vectors
$K_{(a)},\; a=x,y,z,t$: \ba K_{(x)}&=&-\frac{2n(1+C
\cos\theta)\cos\varphi}{\sin\theta}\partial_t
-\sin\varphi\partial_\theta-\cos\varphi\cot\theta\partial_\varphi\,,\nn\\
K_{(y)}&=&-\frac{2n(1+C \cos\theta)\sin\varphi}
{\sin\theta}\partial_t
+\cos\varphi\partial_\theta-\sin\varphi\cot\theta\partial_\varphi\,,\nn\\
K_{(z)}&=&\partial_\varphi+2nC\partial_t\,,\\
K_{(t)}&=& \partial_t\,,\nn \ea the first three forming the algebra
$so(3)$. These formulas extend previously known ones
\cite{misner1,Dowker,Perry} to arbitrary $C$-parameter. Note that
the presence of $\partial_t$-terms in the $so(3)$ subalgebra
reflects the necessity of a compensating time-shift while performing
spatial rotations. The  algebra $so(3)$ can not be integrated to the
group $SO(3)$, but  leads to the unitary representation of the group
$SU(2)$, provided periodic identification of time according to
Misner is performed. This is similar to the case of the magnetic
monopole \cite{Hurst}, for further references see \cite{Dowker}.

The vector potential one-form $A =
\Phi(dt-2n(\cos\theta+C)\,d\varphi)$ is also symmetric under the
action of these four isometries, the Lie derivatives are zero for
all $a$:
$$
{\cal L}_{K_{(a)}}A=0={K_{(a)}}^\nu
A_{\mu;\nu}+K^\nu_{(a);\mu}A_\nu\,.
$$
These symmetries generate four conserved quantities \be
I_a=K_{(a)}^\mu({\dot x}^\nu g_{\mu\nu}+\kappa A_\mu)\,, \ee which
read explicitly: \ba
I_{(x)}&=&2nE (\sin\theta)^{-1}\cos\varphi-\sin\theta {\dot\theta}^2\rho^2-\cot\theta\cos\varphi\left(\rho^2\sin^2\theta {\dot\varphi}+2nE\cos\theta\right) \,,\nn\\
I_{(y)}&=&2nE (\sin\theta)^{-1}\sin\varphi+\cos\theta {\dot\theta}^2\rho^2-\cot\theta\sin\varphi\left(\rho^2\sin^2\theta {\dot\varphi}+2nE\cos\theta\right)\,,\\
I_{(z)}&=&\rho^2 \sin^2\theta {\dot\varphi}+2nE\cos\theta\,,\\
I_{(t)}&=& E\,.\nn \ea They do not depend on the vector potential.

\renewcommand{\theequation}{A.\arabic{equation}}
\setcounter{equation}{0}
\section*{Appendix B: Riemann tensor}
Here we present the Newman-Penrose invariant projections of the
Ricci and Weyl tensor for the metric (\ref{nworm}). Choosing the
null tetrad as
 \ba
l^\mu &=& \frac1{\sqrt{2}}\left(\frac1{\sqrt{f}},\;\sqrt{f},\;0,\;0\right)\,,\nn\\
n^\mu &=&
\frac1{\sqrt{2}}\left(\frac1{\sqrt{f}},\;-\sqrt{f},\;0,\;0\right)\,,\\
m^\mu &=&  \left(\frac{\sqrt{2}n \cot\theta}{\sqrt{r^2+n^2}},\;0,\;
\frac{i}{\sqrt{2}\sqrt{r^2+n^2}},\;
 \frac{1}{\sqrt{2}\sqrt{r^2+n^2}}\right)\,,\nn\\
{\bar m}^\mu &=&  \left(\frac{\sqrt{2}n
\cot\theta}{\sqrt{r^2+n^2}},\;0,\;
\frac{-i}{\sqrt{2}\sqrt{r^2+n^2}},\;
 \frac{1}{\sqrt{2}\sqrt{r^2+n^2}}\right)\,\nn
 \ea
we have one non-zero  projection of the Ricci tensor
 \be
\Phi_{11}=\frac14 R_{\mu\nu}(l^\mu n^\nu + m^\mu {\bar
m}^\nu)=\frac12\frac{e^2}{(r^2+n^2)^2}\,,
 \ee
and one projection of the  Weyl tensor:
 \ba
\Psi_{2}&=&C_{\mu\nu\lambda\tau}l^\mu m^\nu {\bar m}^\lambda n^\tau =
\frac{(m^2+b^2)(r^2-n^2)+n^2r(3m-2r)}{(r^2+n^2)^3}\nn\\
&+&i\,\frac{2nr(m^2+b^2)(r^2+n^2)+nr^4(r-3m)+n^5(r-m)}{(r^2+n^2)^4}
\,.
 \ea
These do not depend on the polar angles. The components of the
Riemann tensor can be read off easily. We give explicitly the ones
used in Sec. 7: \ba {R^r}_{ttr} &=&  -f\dfrac{ \left(
f-1\right) \left( 3r^{2}-n^{2}\right) +4mr}{\left(r^{2}+n^{2}\right)^{2}}\,, \nn\\
{R^r}_{tr\varphi } &=&  {R^r}_{\varphi rt} = -2n\left( \cos \theta
+C\right) f \dfrac{\left( f-1\right) \left( 3r^{2}-n^{2}\right)
+4mr}{\left(r^{2}+n^{2}\right) ^{2}}\,, \nn\\
{R^\theta }_{tt\theta } &=&  {R^\varphi}_{tt\varphi } =
 -f\dfrac{r\left(r-m\right) -\left( r^{2}-n^{2}\right) f}
{\left( r^{2}+n^{2}\right) ^{2}}\,, \nn\\
{R^\theta }_{rr\theta } &=&  {R^\varphi}_{rr\varphi} =
\dfrac{r\left(r-m\right) -\left( r^{2}-n^{2}\right) f}{f\left( r^{2}
+ n^{2}\right) ^{2}}\,, \nn\\
 \ea

\end{document}